\documentclass[twocolumn,twocolappendix]{aastex631}
\usepackage{bm,amssymb,amsmath,times}

\usepackage{newtxtext,newtxmath}

\usepackage[T1]{fontenc}
\usepackage{ae,aecompl}

\usepackage[squaren]{SIunits}	
\usepackage{cuted}      
\usepackage[normalem]{ulem}
\usepackage{longtable}

\makeatletter					

\newcommand{\Msun}{\mathrm{M}_{\sun}}				
\newcommand{\Rsun}{\mathrm{R}_{\sun}}				
\newcommand{\days}{\mathrm{days}}				    
\newcommand{\years}{\mathrm{yrs}}				    

\received{December 12, 2022}
\revised{December 22, 2022}
\accepted{December 23, 2022}
\submitjournal{ApJ}

\shorttitle{SCATTER: A New Common Envelope Formalism}
\shortauthors{Di Stefano et al.}

\begin{document}
\title{SCATTER: A New Common Envelope Formalism}

\author[0000-0003-0972-1376]{Rosanne Di Stefano}
\email{rdistefano@cfa.harvard.edu}
\affiliation{Harvard-Smithsonian Center for Astrophysics, 60 Garden St, Cambridge, MA 02138, US}

\author[0000-0001-9331-0400]{Matthias U. Kruckow}
\affiliation{Yunnan Observatories, Chinese Academy of Sciences, Kunming 650011, China}
\affiliation{Key Laboratory for the Structure and Evolution of Celestial Objects, Chinese Academy of Sciences, Kunming 650011, China}
\affiliation{D\'{e}partement d'Astronomie, Universit\'{e} de Gen\`{e}ve, Chemin Pegasi 51, CH-1290 Versoix, Switzerland}
\affiliation{Gravitational Wave Science Center (GWSC), Universit\'{e} de Gen\`{e}ve, CH-1211 Geneva, Switzerland}

\author[0000-0003-2467-3755]{Yan Gao}
\affiliation{Yunnan Observatories, Chinese Academy of Sciences, Kunming 650011, China}
\affiliation{Key Laboratory for the Structure and Evolution of Celestial Objects, Chinese Academy of Sciences, Kunming 650011, China}
\affiliation{University of Chinese Academy of Sciences}
\affiliation{Institute of Gravitational Wave Astronomy, School of Physics and Astronomy, University of Birmingham, Birmingham, B15 2TT, UK}

\author[0000-0001-5853-6017]{Patrick G. Neunteufel}
\affiliation{Max Planck Institut f\"{u}r Astrophysik, Karl-Schwarzschild-Stra\ss{}e 1, 85748 Garching bei M\"{u}nchen, Germany}

\author[0000-0002-4343-0487]{Chiaki Kobayashi}
\correspondingauthor{Chiaki Kobayashi}
\email{c.kobayashi@herts.ac.uk}
\affiliation{Centre for Astrophysics Research, Department of Physics, Astronomy and Mathematics, University of Hertfordshire, Hatfield, AL10 9AB, UK}

\begin{abstract}
One of the most mysterious astrophysical states is the common envelope (CE) phase of binary evolution, in which two stars are enshrouded by the envelope shed by one of them. Interactions between the stars and the envelope shrinks the orbit. The CE can lead to mergers or to a subsequent phase of interactions. Mergers may involve any combination of two compact objects and/or stars.  Some involving white dwarfs, may produce Type Ia supernovae, while merging neutron stars may yield gamma-ray bursts, and merging compact objects of all kinds produce gravitational radiation.
Since CEs can arise from a variety of different initial
conditions, and due to the complexity of the processes involved, it is
difficult to predict their end states.
 When many systems are being considered, as in population synthesis calculations, conservation principles are generally employed. Here we use angular momentum in a new way to derive a simple expression for the final orbital separation. This method provides advantages for the study of binaries and is particularly well-suited to higher order multiples, now considered to be important in the genesis of potential mergers.
Here we focus on CEs in binaries, and the follow-up paper extends our formalism to multiple star systems within which a CE occurs.
\end{abstract}

\keywords{}

\section{Introduction}
\label{sec:introduction}

The idea of the common envelope (CE) was suggested by \citet{pac76}, to explain the existence of cataclysmic variables
. These are close binaries, typically with orbital periods smaller than about 9 hours, in which a low-mass star transfers mass to a white dwarf (WD). Since the WD must have once been the core of a star far larger than the separation between the WD and its present-day companion, the CE was invoked to bring the WD and its companion star together. There is a natural way for a CE to be generated. When a star fills its Roche lobe (RL) but cannot transfer mass in a stable manner to its companion, its envelope detaches on a dynamical time scale and comes to surround both the core of the donor and its companion. The stars within the envelope spiral closer to each other until the envelope is ejected. 
For a recent review see \citet{2022arXiv221207308R}.

Many physical processes, including gravitation, electromagnetism, and fluid dynamics determine the final state of the system \citep{ijc+13,ijr20}. Advances in simulations are making it possible to explore individual cases in some detail, an approach that already has begun to yield insight into the evolution of the CE phase \citep[e.g., ][]{rt12, pdf+12, sos+20, kso+20}. Hydrodynamical simulations can track the mass in the envelope and incorporate a wealth of physical effects, but can only be applied to a small number of individual systems, and rely on input assumptions that introduce a level of systematic uncertainty that  difficult to quantify.

Much of the interest in the wider community focuses on the end states of the CE phase. The CE can lead to mergers of the components during or after the ejection of the envelope, yielding phenomena as diverse as Type Ia supernovae (SNe~Ia), accretion induced collapse, and double compact object mergers involving neutron stars (NSs) and/or black holes (BHs).  Some of these processes  are important for the origin of elements. Additionally, the end state of the CE can also serve as the starting state for a future phase of mass transfer.

It is therefore tempting proposition to come up with a model that is capable of predicting the end state of a CE evolution phase from its initial conditions, and many efforts have accordingly been directed towards doing so. Among these efforts, some of the most notable have culminated in the $\alpha$ formalism (\citealt{web84}, \citealt{ib93}, see also \citealt{De00}) and the $\gamma$ formalism (\citealt{nvyp00}, \citealt{nt05}, see also \citealt{2004NewA....9..399S}), two prescriptions which are widely used for this purpose (e.g., \citealt{Toonen12}, \citealt{Toonen13}, \citealt{Kor22}). The $\alpha$ formalism has its foundations in conservation of energy, pointing out that the energy used to eject the envelope must have come from the energy donated by the orbit of the binary pair as it shrinks. It asserts that, once the energy conversion efficiency (taking into account the geometry of the system) ${\alpha}_{\rm CE}{\lambda}$ at which orbital energy is converted into envelope internal energy is ascertained, the final state can be easily found. The $\gamma$ formalism, on the other hand, is rooted in conservation of angular momentum, arguing in a similar way that the angular momentum required to expand the envelope must be sourced from the binary orbit. Thus, a given fraction of angular momentum of the entire system must be lost in order to eject a given common envelope which accounts for a given fraction of the total mass of the system, and the ratio between the two fractions is $\gamma$. However, both formalisms run into problems in certain situations, such as when the $\alpha$ formalism encounters situations in which a CE leads to a net increase in the binary separation, and neither is a perfect predictor for all CE end states \citep[see also][]{demarco11,hirai22}. Furthermore, neither prescription can be easily transferred to the study of triple systems and higher-order multiples entirely engulfed by CEs, which are becoming increasingly relevant to studies of stellar evolution today \citep[e.g., ][]{gtl23}.

In this paper we present a new CE formalism which is naturally well suited to triple systems and higher order multiples. In our formalism, each mass in the system transfers angular momentum to material in the CE. Unlike the $\gamma$ formalism, our prescription pays attention to how much each individual component contributes to the angular momentum being taken away, and in doing so, renders the relevant free parameters calculable using first principles. We do not assume anything about the mechanism that ejects the envelope. Instead we focus entirely on the angular momentum transferred from or to the individual stellar components of the multiple-star system. Our primary assumption is that it is the transfer of angular momentum to and from the individual stars that changes their orbits. This allows us to create a mapping between an orbit at the start of the CE and the orbit at the end of the CE phase. The formalism has the nickname SCATTER, for {\sl Single Components' Angular momenTum TransfER}.

The SCATTER approach confers several advantages. First, it predicts changes in the angular momentum of the stellar components and envelope that can be compared with the results of simulations. This type of check can be used both to understand simulation results and to refine the SCATTER formalism. Second, the functional forms we derive are relatively simple, suggesting that they can be successfully applied to a wide range of binary systems from those including WDs to those including BHs. Finally, as mentioned above, the approach we develop generalizes in a direct way to CEs in triples and higher-order multiples. The reason for this is that SCATTER considers the angular momentum transfer from each of the components of a multiple-star system. While the formalism's application to higher-order multiples requires that we develop expressions for the amount of mass primarily exchanging angular momentum with each component, SCATTER provides a robust framework for systems more complex than we explicitly consider here.

We start in \S~\ref{sec:motivation} by discussing the motivation for a new formalism. In \S~\ref{sec:twobody} we introduce our assumptions. We use them to derive a simple expression that maps an initial state, i.e., at the time when the envelope is released by its star, to a final state, i.e., when the binary has settled into its post-CE orbit. This map involves free parameters.  \S~\ref{sec:parameters} is devoted to the enterprise of deriving physically and observationally reasonable parameter values, while in \S~5 we find a formula that allows the key parameter to be selected in a manner that is consistent with all the data we have and which is easy to apply. In \S~\ref{sec:applications} we apply our formalism to randomly generated binaries, each of which includes a WD. We compare the results with those derived using standard implementations of the ``$\alpha$'' and ``$\gamma$'' formalisms in \S~\ref{sec:comparison}. Section~\ref{sec:conclusions} is devoted to an overview of the work.

\section{Motivation}
\label{sec:motivation}

Consider a binary whose components are close enough to each other that the first-evolved star fills its RL after its core has a mass greater than about $0.1\,\Msun$. If the mass transfer is unstable on a dynamical time scale, the donor's envelope becomes a CE, encompassing the core and its companion star. The core and companion may merge. If, however, the orbit stops shrinking, its orbital separation at the end of the CE phase, $a(f)$, is generally small enough that the companion, Star~2, will eventually fill its RL. Should Star~2 have a mass larger than that of the remnant of Star~1, and/or be at least somewhat evolved, the second phase of mass transfer initiated by the RL filling of Star~2 may also yield a CE. Many interacting binaries thus experience two CE phases. Evolution with two CEs is particularly common in binaries whose final fate\footnote{The merger may occur within the CE or it may occur at a later time, after dissipative forces such as the emission of gravitational radiation have had time to further shrink the orbit.} is the merger of two compact objects \citep[e.g., ][]{Wu18,mar21}. 

The prevalence of CE evolution suggests that uncertainties in the CE phase lead to uncertainties in the rates, times, and characteristics of many events currently of great interest, such as the mergers of BHs, short gamma-ray bursts, or the production of SNe~Ia \citep[e.g.,][]{iben84,tut93}.
Such events are important: BH and/or NS mergers provide direct insight into the properties of BHs and NSs.
SNe~Ia light up distant parts of the universe, allowing us to study the accelerated expansion of the Universe.
SNe~Ia also influence the host galaxies, for example the chemical evolution of galaxies (GCE), which has been used for constraining the formation and evolution of galaxies in the universe.
Binary population synthesis (BPS) is motivated by the goal of computing the number of events (such as mergers, or SNe~Ia), the time the events occur relative to star formation, and the characteristics of the events (e.g., \citealt{bel02,rui09,men10,men14,ktl+18,han20,man21,bri22}; see also \citealt{toonen14,mb22} for a comparison). To do this, a set of calculations starts with a population of binaries, and evolves each to determine whether it becomes one of the systems of highest interest. Because the CE is an evolutionary phase through which many of the binaries pass, but which is notoriously difficult to constrain, it often introduces a substantial uncertainty into BPS results. 

Consider for example, the progenitors of SNe~Ia. The nature or natures of SNe~Ia progenitors has been debated for more than half a century, but still there is no perfect progenitor model. The two mainstream models can be described as follows. In the {\sl double-degenerate} (DD) model, two WDs in a close orbit eventually merge and if the mass of the merger product reaches a critical mass, often considered to be the Chandrasekhar mass (about $1.4\, \Msun$) an SN~Ia ensues \citep{neunteufel20}. In {\sl single degenerate} (SD) progenitor binaries, a WD has an extended companion whose donated mass can bring the WD to the point of explosion. Both models are incorporated into BPS simulations. See \citet{jha19,soker19,ruiter20} for a review including the other scenarios \citep[e.g., ][]{raskin09,rosswog09,rui11,Foley13,McCully14,soker15,shen18}. Unfortunately, BPS calculations have so far been unable to satisfactorily reproduce population properties such as delay times and chemical enrichment of galaxies, see Figure~1 (see also \citealt{kob22} for the details). Since $\alpha$ elements (e.g., O) are mainly produced from core-collapse supernovae on a short timescale ($\mega\years$) and the majority of Fe is produced from SNe~Ia on a longer timescale ($\gtrsim \unit{0.1}{\giga\years}$), the [$\alpha$/Fe] ratio has been used as a proxy of ages of stellar populations in galaxies. Although there are successful analytical models for including SNe~Ia in GCE \citep{greggio83,matteucci86,ktn+98,gre05,kob09}, there are no BPS models that can reproduce the observations in the Milky Way galaxy \citep[obtained from high-resolution spectra of nearby stars][]{zmy+16}, namely, the evolution of elemental abundance ratios in the solar neighbourhood.
The SN~Ia delay times are also important for constraining the star formation histories from the observed [$\alpha$/Fe] ratios of early-type galaxies.
\begin{figure}
    \centering
    \includegraphics[width=\columnwidth]{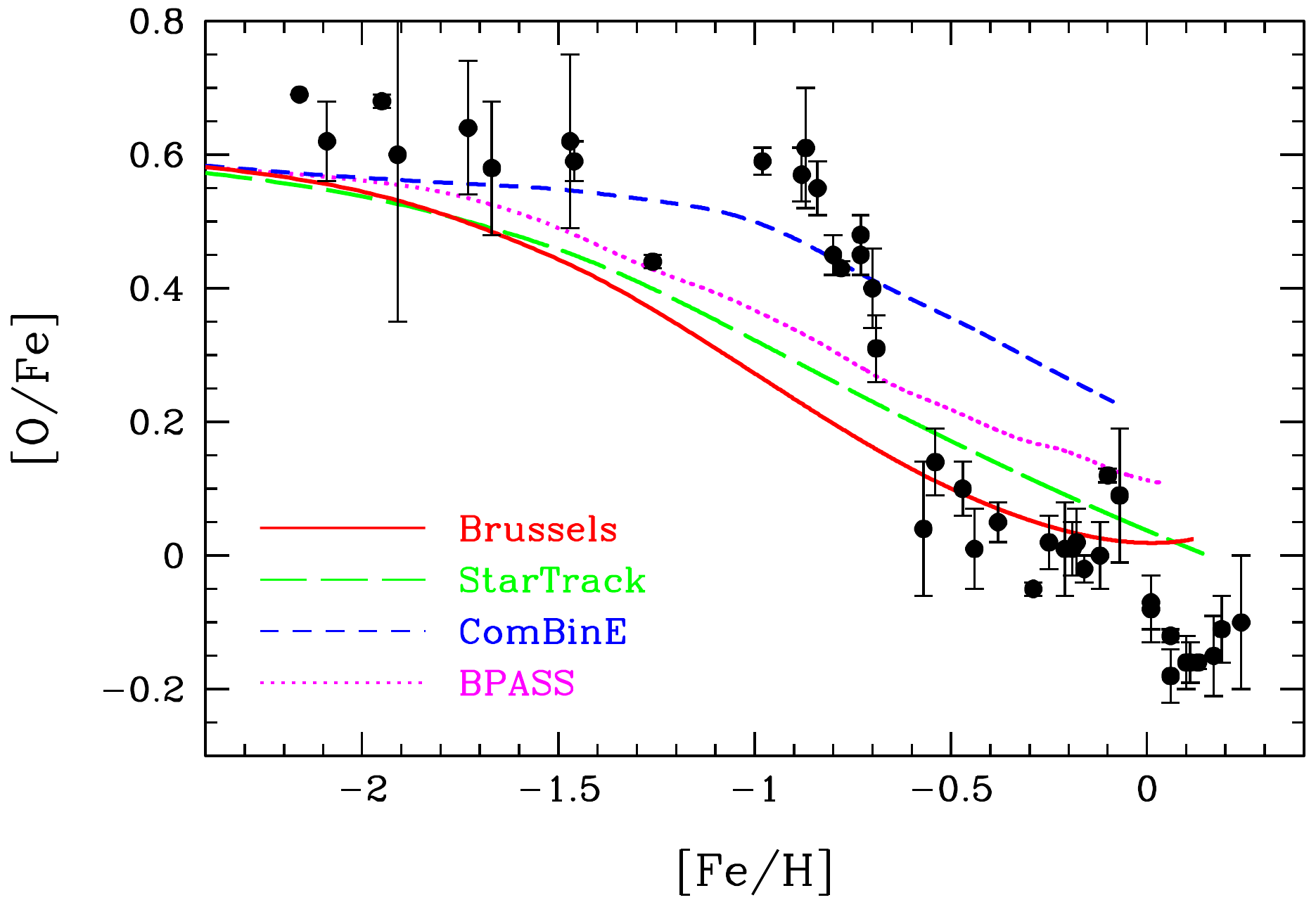}
    \caption{Comparison between observational data (black symbols) and specific implementations of four BPS models.  The red solid line uses the Brussels model \citep{men10,men12}. The green long-dashed line uses StarTrack \citep{rui09,rbs+14}. The blue short-dashed line uses ComBinE (private communication based on \citealt{ktl+18}). The magenta dotted line uses BPASS \citep{bri22}.}
\end{figure}

The uncertainties in the CE parameters are also responsible for significant uncertainties in the rates of double BH mergers \citep[e.g., ][]{bb98, dbf+12, cbk+18, ktl+18, vns+18, mg18, tsy+21, bfz+21, obi21}, for more details see e.g. Figures 3.47, B.48, and B.49 in \citet{kru18} and a recent summary in \citet{mb22}. 
This is because gravitational wave emission, hence the time between star formation and merger,   depend strongly on the orbital separation  \citep{pet64}. Hence, the results of CE evolution have  major impacts on the rates of such merging events. 
For mergers involving NSs the uncertainties of CE evolution are smaller but still important, while kicks during NS formation can introduce significant eccentricity and also influence on the rate of mergers that include NSs.

\section{The Formalism}
\label{sec:twobody}

\subsection{Basic assumptions}
\label{sec:assumptions}
The SCATTER formalism aims to derive the final (post-CE) separation between the binary components, and does not rely on understanding the mechanisms ejecting the CE. The physics of CE evolution is complex.  Both energy transfer and
angular momentum transfer almost certainly each take place through
a variety of mechanisms, and at a range of times from the start of the
CE.  For example, angular momentum transfer to the envelope may occur at
early stages, before the companion enters the CE or when it is still
in the outer envelope \citep{2004NewA....9..399S}. The SCATTER formalism makes no
assumptions about the timing or mechanisms responsible for the transfer of
angular momentum, or about the state of the CE and its reaction to
interactions with the binary's components.
We make only the following basic physical assumptions.

\begin{itemize}
    \item The initial state at the start of the CE is defined by the angular momentum of the binary, which is the sum of the angular momenta of its components. The same is true of the final state. Thus, the  change between the initial and final angular momentum of the binary's components  provides a map from the binary's initial state  to its final state. We assume that the angular momentum lost by each of the binary's components during the CE was transferred from it to the envelope. The effect of the stars on the envelope, and the mechanisms leading to its ejection are not explicitly considered. 

    \item Angular momentum is conserved between the initial and final state of the binary. That is,  angular momentum lost by the binary is transmitted to the envelope. The angular momentum transferred from each component of the binary can be viewed as providing angular momentum to a quantitatively well-defined fraction of the envelope. We do not attempt to identify the specific portions of the envelope that receive angular momentum from each star. We do estimate the quantity of mass that received angular momentum primarily from each component star.

    \item We take the initial state to be that in which the envelope has just been released. The mass of the donor's core and its companion are $M_\mathrm{c}$ and $M_\mathrm{2}$, respectively. The orbital radius is $a(0).$ The envelope mass is $M_\mathrm{e}$. Star~1 is the donor, with pre-CE mass  $M_1=M_\mathrm{e}+M_\mathrm{c}$.
    
    \item In this paper we do not consider mass exchange. In other words, the masses of the two objects spiraling toward each other are constant. This restriction can be lifted within the formalism by incorporating mass exchange in Equation~\eqref{eq:dLL}. 
    
    \item The initial and final orbits of the binary are circular. 
    
\end{itemize}

We also note that this formalism can be applied to take advantage of the fact that the angular momentum is a three vector. In this paper, however, we consider just a single component of the angular momentum sufficient to describe a binary orbit.
In addition, the formalism can be naturally extended to predict end states of CEs that occur in higher order multiples. This is the subject of a separate paper.

\subsection{Derivation}
\label{sec:derivation}
The CE starts at the time the donor star, Star~1 with mass $M_\mathrm{1}=M_\mathrm{e}+M_\mathrm{c}$, fills its RL. At the moment of RL filling, the core mass of Star~1 is $M_\mathrm{c}$ and its envelope mass is $M_\mathrm{e}$. The orbital separation between the stellar centers is $a(0)$. On a dynamical time scale Star~1 loses its envelope. The stripped core and Star~2 spiral toward each other within the envelope.  Whatever the time scale of the spiral-in or plunge-in, the CE itself takes some time to dissipate. \citet{ijc+13} pointed out that roughly one in five planetary nebulae may correspond to CEs. At typical expansion velocities in the range of tens to a few hundred km~s$^{-1}$, it takes $10^4$~yr to more than $10^5$~yr for the envelope to disperse.  Our goal is to determine whether the two masses merge while still inside the CE and, if they don't, to find a physically reasonable expression for their final orbital separation, $a(f)$.

\subsubsection{\texorpdfstring{$\Delta M_{\mathrm{e},i}$}{Portion of the envelope interacting with star i}}
\label{sec:massassociation}

All CE formalisms employ a fundamental principle to relate the final state to the initial state. We use angular momentum conservation, assuming that the evolution of the orbit can be computed by calculating the orbital angular momentum each star imparts to some mass in the CE. Thus, Star~$i$ exchanges angular momentum with an amount of envelope mass $\Delta M_{\mathrm{e},i}$. Note that $\Delta M_{\mathrm{e},i}$ refers to mass in the CE. The subscript $i$ simply indicates that this amount of mass exchanged angular momentum mainly with Star~$i$. The following equation applies.
\begin{equation}
    \sum_{i} \Delta M_{\mathrm{e},i} = M_\mathrm{e},
\end{equation}
where the sum covers all of the stellar components. For binaries, $i\in\{\mathrm{c},2\}$. This approach treats the CE as a union of distinct sets, each of which draws angular momentum from a single star. This  may suggest that the envelope can be neatly divided into contiguous regions in which one stellar component is primarily responsible for angular momentum transmission. In fact, the situation can be highly complex, especially because the positions and velocities of small masses in the envelope change in time. Furthermore, bits of mass in the envelope interact with other bits of envelope mass. In our formalism we assume that each small bit of envelope mass that has gained angular momentum during the CE phase, drained angular momentum from the binary's components and that it is possible, perhaps only in a probabilistic sense, to say that one of the binary's component masses was the primary donor of angular momentum to that bit of mass.  

\begin{figure}
\begin{center}
    \includegraphics[width=\columnwidth]{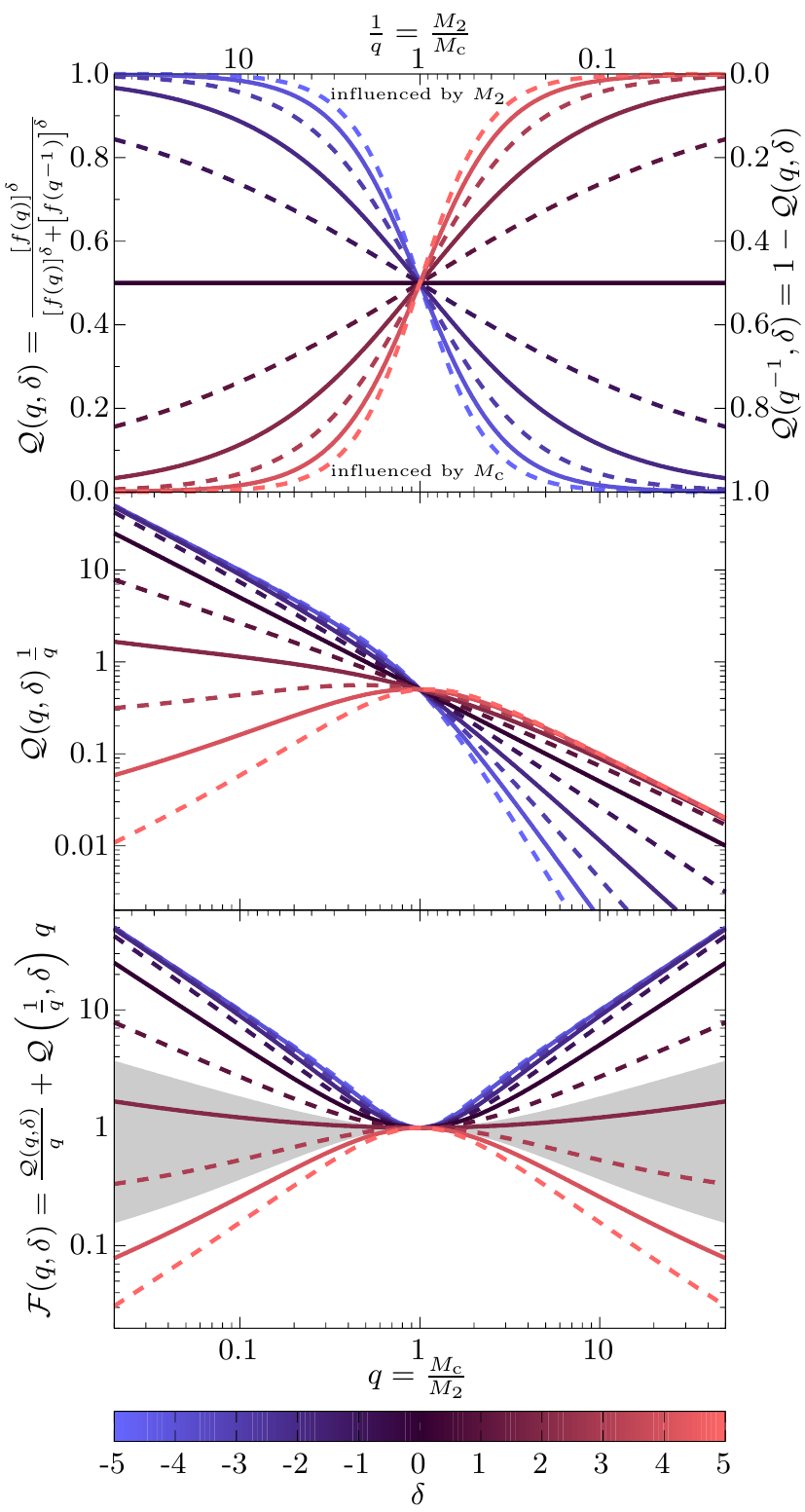}
    \caption{The influence of the choice of $\delta$. Top: $\mathcal{Q}$ for different $\delta$s, which is colour coded (lines of even numbers of $\delta$ are solid and odd numbers are dashed). The part below the lines is the amount of mass influenced by $M_\mathrm{c}$ (left y-axis) and above is the amount of mass influenced by $M_\mathrm{2}$ (right y-axis). The plot is point symmetric at $q=M_\mathrm{c}/M_\mathrm{2}=1$ and $\mathcal{Q}(1,\delta)=0.5$. Middle: The angular momentum change caused by $M_\mathrm{c}$ for the same functions as in the top panel. Bottom: The sum of the angular momentum change by both stars, $\mathcal{F}\left(M_\mathrm{c}/M_\mathrm{2},\delta\right)$. The gray area marks $1.5 \leq \delta \leq 3.5$.}
    \label{fig:qQ}
\end{center}
\end{figure}

We do not attempt to identify the source of angular momentum for each part of the envelope or even to assign a probability of angular momentum transfer to each of the binary components. Instead, we employ a simple expression that emerges from the RL formalism to estimate the quantity, $\Delta M_\mathrm{e,c}$ of the CE mass that drains orbital angular momentum primarily from $M_\mathrm{c}$. Similarly, we estimate $\Delta M_\mathrm{e,2}$, which drains orbital angular momentum primarily from $M_\mathrm{2}$. Both mass and rotation play roles in determining the effective potential. It is therefore useful to consider the RL of each star. If $q\equiv M_\mathrm{c}/M_\mathrm{2}$, then the effective radius of the core's RL is $f(q)\, a$, where $a$ is the separation between Star~2 and Star~c, and 
\begin{equation}
    \label{eq:roche}
    f(q)=\frac{0.49\, q^{\frac{2}{3}}}{0.6\, q^{\frac{2}{3}} + \ln\left(1+q^{\frac{1}{3}}\right)},
\end{equation}
this function is called $r_\mathrm{L}$ in \citet{egg83}.

We define
\begin{equation}
    \label{eq:Q}
    \mathcal{Q}\left(q,\delta\right) \equiv \frac{\left[f\left(q\right)\right]^\delta} {\left[f\left(q\right)\right]^\delta + \left[f\left(q^{-1}\right)\right]^\delta}.
\end{equation}
The parameter $\delta$ corresponds to a dimension, so that roughly speaking, $[a\,f(q)]^\delta$ corresponds to a $\delta$-dimensional volume.

Note that $\mathcal{Q}(q,\delta)+\mathcal{Q}(q^{-1},\delta)=1$. We therefore define 
\begin{equation}
    \Delta M_\mathrm{e,c} \equiv \mathcal{Q}(q,\delta)\, M_\mathrm{e},
\end{equation}
to be the fraction of the CE mass draining orbital angular momentum primarily from $M_\mathrm{c}$, with an analogous expression for $\Delta M_\mathrm{e,2} = \mathcal{Q}(q^{-1},\delta)\, M_\mathrm{e}$.

The functional form of $\mathcal{Q}$ has the advantages of (1)~assuring that the more massive star interacts with a larger portion of the CE; (2)~having a symmetric form; and (3)~being normalized. The applicability of this function can be tested through comparisons with observations. Future studies may alter the definition of $\mathcal{Q}$, perhaps to better represent observational constraints.

\subsubsection{Angular Momentum}
\label{sec:angularmomentum}
The orbital angular momentum may be written as 
\begin{equation}
    \label{eq:L}
    L = M_\mathrm{1}\, M_\mathrm{2}\, \sqrt{\frac{G\, a}{M_\mathrm{b}}},
\end{equation}
where $M_\mathrm{b}$ refers to the total mass of the binary at the start of the CE: $M_\mathrm{b} \equiv M_\mathrm{1}+M_\mathrm{2}$, where $M_\mathrm{1} = M_\mathrm{c}+M_\mathrm{e}$, $G$ is the gravitational constant, and $a$ is the current semi-major axis of a circular orbit. As the CE starts,  $a=a(0)$. During the CE phase, the value of the $M_\mathrm{c}$ stays the same, but the envelope is lost, effectively decreasing the total mass $M_\mathrm{b}$ of the binary as well as the mass of Star~1. The orbital separation also shrinks. The expression for orbital angular momentum, $L$, leads to the following equation.
\begin{equation}
    \label{eq:dLL}
    \frac{\mathrm{d}L}{L}=\frac{1}{2}\, \frac{\mathrm{d}a}{a} - \frac{1}{2}\, \frac{\mathrm{d}M_\mathrm{b}}{M_\mathrm{b}} + \frac{\mathrm{d}M_\mathrm{1}}{M_\mathrm{1}} + \frac{\mathrm{d}M_\mathrm{2}}{M_\mathrm{2}}.
\end{equation}

This expression is completely general.  If the three-dimensionality of the angular momentum is important, we have three such equations, one for each 
spatial direction.  Here we make the simplification of treating the angular momentum as a scalar quantity, which is equivalent to assuming the the vector normal to the orbital plane points in the same direction at all times. 

If known external torques act on the binary, we can compute the quantity $dL/L$. 
To find an expression for $dL/L$ we note that the CE exerts a torque on each component of the binary. Each binary component  mass therefore experiences a change in its angular momentum.
For the core, we have
\begin{equation}
    \label{eq:dLLc}
    \frac{dL_\mathrm{c}}{L}\, = \eta_\mathrm{c}\, Q\left(\frac{M_\mathrm{c}}{M_\mathrm{2}}, \delta\right)\, \frac{dM_\mathrm{e}}{M_\mathrm{c}}\, \frac{M_\mathrm{2}}{M_\mathrm{c}+M_\mathrm{2}} .
\end{equation}
This equation defines a quantity $\eta_\mathrm{c}$: the fractional change in $L_\mathrm{c}$ is equal to the specific angular momentum of the core times the amount of envelope mass with which it interacts, times the constant of proportionality $\eta_\mathrm{c}$.

For Star~2 we have
\begin{equation}
    \label{eq:dLL2}
    \frac{dL_\mathrm{2}}{L}\, = \eta_\mathrm{2}\, Q\left(\frac{M_\mathrm{2}}{M_\mathrm{c}}, \delta\right)\, \frac{dM_\mathrm{e}}{M_\mathrm{2}}\, \frac{M_\mathrm{c}}{M_\mathrm{c}+M_\mathrm{2}} .
\end{equation}
 Thus, the quantity on the left-hand side of equation~(\ref{eq:dLL}) can be expressed as follows:
\begin{equation}
    \frac{\mathrm{d}L}{L}=\frac{\mathrm{d}L_c}{L}+\frac{\mathrm{d}L_2}{L} .
\end{equation}

With an explicit expression for $dL/L$, we can plug into equation~(\ref{eq:dLL}) 
 derive the functional form of $da/a$.
 \begin{equation}
    \frac{\mathrm{d}a}{a}=2\, \Bigg[\frac{\mathrm{d}L_c}{L}+\frac{\mathrm{d}L_2}{L} \Bigg] 
    + \frac{dM_\mathrm{b}}{M_\mathrm{b}} -2\, \frac{dM_1}{M_1} -2\, \frac{dM_2}{M_2} .
 \end{equation}
 
Integration from the initial to final masses then yields $a(f)/a(0)$ or its inverse.  
In this integration we 
assume that Star~2 does not gain mass during the CE. This assumption can easily be relaxed be relaxed.
When, however, there is  no significant accretion onto the companion, then $\int \mathrm{d}M_\mathrm{2}\approx 0$. 
Similarly, $M_\mathrm{c}$ remains constant.
The binary loses the mass $M_\mathrm{e}$, the same as the mass lost by Star~1, thus $\int \mathrm{d}M_\mathrm{1} \approx -M_\mathrm{e} \approx \int \mathrm{d}M_\mathrm{b}$.

This procedure yields
\begin{equation}
\begin{aligned}
    \label{eq:ar}
    \left(\frac{a(0)}{a(f)}\right)\, \left(1+\frac{M_\mathrm{e}}{M_\mathrm{c}}\right)^2\, \left(1+\frac{M_\mathrm{e}}{M_\mathrm{c}+M_\mathrm{2}}\right)^{-1} = \\
    \exp\left(\left[\frac{2\, M_\mathrm{e}}{M_\mathrm{c}+M_\mathrm{2}}\right]\, \left[\eta_\mathrm{c}\, \frac{M_\mathrm{2}}{M_\mathrm{c}}\, Q\left(\frac{M_\mathrm{c}}{M_\mathrm{2}},\delta\right) + \eta_\mathrm{2}\, \frac{M_\mathrm{c}}{M_\mathrm{2}}\, Q\left(\frac{M_\mathrm{2}}{M_\mathrm{c}},\delta\right) \right]\right).
\end{aligned}
\end{equation}

The ratio of the initial to final orbital separations depends only on mass ratios: $q = M_\mathrm{c}/M_\mathrm{2}$, and its inverse, and $q_\mathrm{ec} \equiv M_\mathrm{e}/M_\mathrm{c}$. Note that the $M_\mathrm{e}/(M_\mathrm{c}+M_\mathrm{2})$ can be written as 
\begin{equation}
     \frac{M_\mathrm{e}}{M_\mathrm{c}+M_\mathrm{2}}
     = q_\mathrm{ec}\, \frac{q}{1+q}.
\end{equation}
Dependence on only mass ratios suggests applicability to a wide range of systems across mass ranges. The specific value of the donor's starting stellar mass comes into play only through the value of $a(0)$, since
\begin{equation}
\label{eq:a0}
    a(0) = \frac{R_\mathrm{1}(0)}{f\left(\frac{M_\mathrm{1}}{M_\mathrm{2}}\right)},
\end{equation}
where $R_\mathrm{1}(0)$ and $M_\mathrm{1}$ are, respectively, the radius and mass of the donor star at the time of RL filling.

A convenient way to write Equation~\eqref{eq:ar} is as follows:
\begin{equation}
\label{eq:af}
\begin{aligned}
    a(f) = \\
    a(0)\, \left[\frac{(1+q)\, (1+q_\mathrm{ec})^2}{1+q\, (1+q_\mathrm{ec})}\right]\, \exp\left[-2\, \eta\, \left(\frac{q\, q_\mathrm{ec}}{1+q}\right)\, \mathcal{F}(q,\delta)\right],
\end{aligned}
\end{equation}
where
\begin{equation}
    \label{eq:F}
    \mathcal{F}\left(q,\delta\right) = \mathcal{Q}\left(q,\delta\right)\, \frac{1}{q} + \mathcal{Q}\left(\frac{1}{q},\delta\right)\, q.
\end{equation}
Here we simplified the equation~\eqref{eq:F} by assuming that the proportionality parameter is the same for both stellar components, $\eta=\eta_\mathrm{c}=\eta_\mathrm{2}$.

By itself, $Q(q,\delta)$ is a measure of only how the mass ratio influences the amount of envelope mass  accepting orbital angular momentum from each component of the binary. The function $\mathcal{F}$ is a sum of terms, one of which is illustrated in the middle panel. The combined effects of the two terms forming $\mathcal{F}$ has a kind of self regulation effect, as shown in the bottom panel of Figure~\ref{fig:qQ}. Nevertheless, across values of the mass ratio, $\mathcal{F}$ can achieve a wide range of values depending on $\delta$.  Extreme values of $\delta$ could therefore produce extreme values of $a(f),$  unless we were to invoke extreme values of $\eta.$  On the other hand, for values of $\delta$ in the vicinity of $2$, there is little variation on the value of $\mathcal{F}$ as $q$ varies over almost $4$ orders of magnitude. Whatever the value of $\delta$, there is an interval around $q=1$ for which the value of $\mathcal{F}$ is unity; the size of this region is also smaller for large values of $\delta$.

\subsection{Generalizations}
The derivation above can be extended in several ways.  One way would be to make different choices for the functions that describe quantities such as the amount of CE mass to which each binary component transfers angular momentum. Another function could, for example, be used in place of the function $Q(q,\delta)$.  Simulations of CEs may suggest alternatives. If the functional form changes, the parameters employed in the function would likely be differed as well. Thus, the basic ideas of SCATTER as outlined in \S 3.1 can be implemented in different ways.  We note that the choices we have made are physically motivated.  Furthermore, we show in \S 4 that they provide a consistent set of predictions for large sets of known post CE binaries. 

The second type of extension can be carried out using our functional choices, or using others. These include, for example,  (1)~allowing
the components to gain or lose mass during the CE; (2)~treating the angular momentum as a vector, which would have the physical effect of introducing spin to the post CE binaries; (3)~considering systems with higher multiplicity; (4) using post-CE binaries of types different from those we employ in the later sections of the paper. With regard to the latter point, such systems could include planetary systems that have experienced CEs, and also  systems in which the compact objects are BHs and NSs, without a WD component.

\section{The Parameters}
\label{sec:parameters}

Our implementation of the SCATTER formalism includes three parameters: $\delta$, $\eta_\mathrm{c}$, and $\eta_\mathrm{2}$. Given values of these parameters, we could start with any initial CE state and derive the final orbital separation.
In this section we discuss the roles of the parameters within the formalism and try to gain insight from observations of CE final states in order to estimate values of those parameters.

For now we consider that both binary components transmit a similar proportion of their specific angular momentum to the CE. Thus: $\eta=\eta_\mathrm{c}=\eta_\mathrm{2}$. This leave $\eta$ and $\delta$ as the to-be-determined quantities.  
The parameter $\eta$ has a physical meaning, because $\eta$ is the parameter which multiples the specific angular momentum imparted to mass in the envelope. The components of the binary impart a larger portion of their specific angular momentum to the envelope for larger values of $\eta$. We may think of it as a sort of inverse efficiency parameter. Given this, both very large and very small values of $\eta$ should be excluded. We might expect that roughly $0.1 < \eta < 10$.  Even employing the simple assumption that $\eta$ is the same for both binary components, it is natural to anticipate that its value depends on physical characteristics of the components, in particular their mass ratio. The value of $\eta$ is also expected to depend on the properties of the envelope, in particular the envelope mass relative to the masses of the components.
The form of Equation~\eqref{eq:af} then tells us that, if $\eta=\eta(q,q_\mathrm{ec})$, then the value of the final orbital separation depends only on $q$ and $q_\mathrm{ec}$. In the absence of magnetic fields or external torques, the spin-down of the binary is governed by just these two variables.

The value of $\delta$ determines how much of the envelope's mass draws angular momentum from each component of the binary. It may be thought of as a dimensionality. Equation~\eqref{eq:Q} shows that the larger the value of $\delta$, the larger the value of $Q$ associated with the most massive binary component (i.e., $f(q)$ increases with the mass ratio, $q$). The dependence of $Q$ on $q$ and $\delta$ is illustrated in the top panel of Figure~\ref{fig:qQ}. We have considered values of $\delta$ extending from $-5$ to $5$. The rough correspondence with a dimensionality suggests that the value of $\delta$ should be positive. Although there is not an exact correspondence between the amount of mass influenced and a volume whose size is determined by $\delta$, we might expect values of $\delta$ to be near $3$ and will be discussed in more detail in \S~\ref{sec:derivingdelta}.

\subsection{Post-CE Binaries}
\label{sec:PCEB}
\begin{table}
    \caption{\label{tab:obsdata}An overview of the systems selected from \citet{knd+21}. For each type of binary the number of the systems and the ranges for the main parameters (period and masses of the two stars) are quoted.}
    \centering
    \small
    \begin{tabular}{ccccc}
    \hline
        Type & \# & Period & \!Mass of WD\! & \!\!Mass of Comp.\!\! \\
         & & $[\days]$ & $[\Msun]$ & $[\Msun]$\\
        \hline
        WD+MS & $\!187\!$ & $\!0.0651-88.1805\!$ & $\!0.07-1.23\phantom{0}\!$ & $\!0.087-3.4\phantom{0}\!$ \\
        WD+NS & $\!\phantom{0}28\!$ & $\!0.1024-95.1741\!$ & $\!0.16-1.37\phantom{0}\!$ & $\!1.14\phantom{0}-3.4\phantom{0}\!$ \\
        WD+WD & $\!\phantom{0}34\!$ & $\!0.0048-\phantom{0}2.21\phantom{00}\!$ & $\!0.15-0.813\!$ & $\!0.167-1.06\!$ \\
        \hline
    \end{tabular}
\end{table}

We use the database of hundreds of post-CE binary  candidates compiled in \citet{knd+21}. This catalog summarizes several decades of literature about the observation of short period binaries which are potential end states of CE evolution. The most useful information, including mass estimates, originates from observations of eclipsing binaries and/or spectroscopic observations.
In \citet{knd+21} the candidates for post-CE binaries are selected by their present-day short orbital period (below $\unit{100}{\days}$). Here we require that at least one component be a WD, and that there are mass estimates for each of the post-CE binary's components.  Hence we exclude all flagged systems\footnote{Most flags indicate that some of the data is assumed instead of measured or the binary has evolved significantly since the ejection of the CE.} and systems with only lower or upper mass limits.
In Table~\ref{tab:obsdata} we show the properties of the post-CE binaries we utilize in this work.

\subsection{Input from Stellar and Binary Evolution}
\label{sec:stellarevolution}

To use the post-CE binaries to create a map between CE initial and final states, we invoke
results from both stellar and binary evolution.  
Stellar evolution relates the donor's radius to its evolutionary state. We consider only donors whose cores will become WDs.
In these cases, the simplest prescription for the donor's radius, $R_1(0)$ depends primarily on $M_\mathrm{c}$
\citep{1987ApJ...319..180J}. 
At $t(0)$, the moment of RL filling, when the core mass is $M_\mathrm{c}$, $M_\mathrm{1}$ may be close in value to the donor's zero-age main sequence (ZAMS) mass if there have not yet been significant winds, but it may also be significantly smaller. In either case, the mass of the envelope is $M_\mathrm{e}=M_\mathrm{1}-M_\mathrm{c}$. Furthermore, when the core mass is larger than roughly $0.2\, M_\odot$, the stellar radius has only a weak dependence on the initial mass of the donor. We therefore select possible donor initial masses 
to be approximately equal to $M_\mathrm{e}$. 
 While detailed evolutionary calculations would refine these estimates,  a simple unified approach allows us to focus on the effects of CE evolution. 

At the time of Roche-lobe filling, the orbit is circular, with radius given by the equation~\eqref{eq:a0}.
Thus, if, in addition to having post-CE masses, we also knew the value of $M_\mathrm{e}$, we could use the equation~(\ref{eq:af}) to compute $a(f).$

To select a possible value (or values) of $M_\mathrm{e}$, we used stellar models to compute both its minimum and maximum possible values (Appendix A). We estimated the minimum envelope mass to be $(0.9\, \Msun - 0.85\, M_\mathrm{c})$. If the result was an envelope less massive than $0.6\, \Msun$, we instead set $M_\mathrm{e}^\mathrm{min}$ to $0.6\, \Msun$.  For WDs with mass greater than $0.55\, \Msun,$ we took the $M_\mathrm{e}^\mathrm{min}$ to be $(0.8\, \Msun +2.1\, M_\mathrm{c})$. For all WD's we took the maximum envelop mass, $M_\mathrm{e}^\mathrm{max}$ to be $(1.3\, \Msun +10.5\, M_\mathrm{c})$. If, however, $M_\mathrm{e}^\mathrm{max}$ was larger than $8\, \Msun$, we instead set it to $8\, \Msun$. 

The effective radius of the donor's  RL at the start of the CE is $R_\mathrm{RL}(0)=a(0)\times f(\frac{M_\mathrm{e}+M_\mathrm{c}}{M_\mathrm{2}})$. We can also define a ``final'' effective RL radius to be
$R_\mathrm{RL}(f)=a(f) \times f(\frac{M_\mathrm{c}}{M_\mathrm{2}})$. This allows us to introduce a variable
\begin{equation}
\label{eq:rd}
{\cal R}_\mathrm{d}\equiv\frac{R_\mathrm{RL}(0)}{R_\mathrm{RL}(f)} = \frac{a(0)}{a(f)}\times \frac{f(\frac{M_\mathrm{e}+M_\mathrm{c}}{M_\mathrm{2}})}{f(\frac{M_\mathrm{c}}{M_\mathrm{2}})}.
\end{equation}
Large values of ${\cal R}_\mathrm{d}$ correspond to cases in which the CE phase significantly shrunk the binary's orbit.
When we produce graphs of post-CE binaries, we mark in green those with ${\cal R}_\mathrm{d} > 100;$ these are almost certainly CE end states. Binaries with $100>{\cal R}_\mathrm{d} >20$ are probable CE end states, and are shown as blue points.
This with ${\cal R}_\mathrm{d} <20$ are red, and may not represent true CE end states.

At time $t(f),$ when the CE phase ends, the orbital separation is $a(f)$. The orbit generally continues to evolve. If, for example, the post-CE binary comprised of compact objects and if it is small enough that gravitational radiation carries away significant angular momentum, the binary continues to shrink. 
The time to merger can then be estimated to be\footnote{We assume that $a(f)$ is much larger than the separation at which one of the post-CE binary's
components will fill its RL.}
\begin{equation}
\label{eq:tmerge}
    \tau_\mathrm{merge} = \unit{1.5\times 10^8}{\years} \left[\frac{a(f)}{\Rsun}\right]^4 \left[\frac{\Msun^3}{M_\mathrm{c}\, M_\mathrm{2}\, M_\mathrm{t}}\right] ,
\end{equation}
where the total mass of the binary system at the end of CE is defined as $M_\mathrm{t}\equiv M_\mathrm{c}+M_\mathrm{2}$.
If the orbit is small but the WD's companion is a main sequence (MS) star, magnetic braking (rather than gravitational radiation) may bring the binary's components closer together, and an epoch of mass transfer can ensue. If the mass transfer is stable, the system will eventually become a cataclysmic variable, in which a WD receives mass at a low rate from a close companion. In this case, the MS star fills its RL. In the figure of Section \ref{sec:WDMS}, if the value of $a(f)$ in the post-CE state is such that the MS star is close to Roche-lobe filling, we mark the
system with a special symbol, a large gold pentagon, to indicate that the post-CE binary may have evolved, so that the binary separation we measure is not the same as the CE end state.

\subsection{Constraining \texorpdfstring{$\delta$}{delta}}
\label{sec:derivingdelta}

Our first step is to estimate the value of $\delta$, which appears only in the factor $\mathcal{F}(q,\delta)$. The bottom panel of Figure~\ref{fig:qQ} plots the function $\mathcal{F}(q,\delta)$ as a function of $q$ for the full range of $q$ we need to consider. Each curve employs a single value of $\delta$, ranging from -5 [dashed, blue top-most curve] to 5 [dashed red bottom-most curve]. Note that, for both the smallest and largest values of $\delta$, the value of $\mathcal{F}(q, \delta)$ varies over more than an order of magnitude. This suggests that $\eta$ would also be highly variable if $\delta$ is either very large or small. Since $\delta$ defines the dimensionality of the volume of matter interacting with the binary components, it seems likely that its value should lie between 1 and 4. The gray area in the bottom panel of Figure~\ref{fig:qQ} shows the region in which $\delta$ ranges between 1.5 and 3.5. The solid (dashed) line in that region corresponds to $\delta=2$ ($\delta=3$). For values of $\delta$ in this more limited range, the variation in $\eta$ from binary to binary would not have to be large because the variation of $\mathcal{F}(q,\delta)$ is modest over the range of likely $q$ values.

\begin{figure}
    \centering
    \includegraphics[width=\columnwidth, clip, trim=30px 400px 30px 100px]{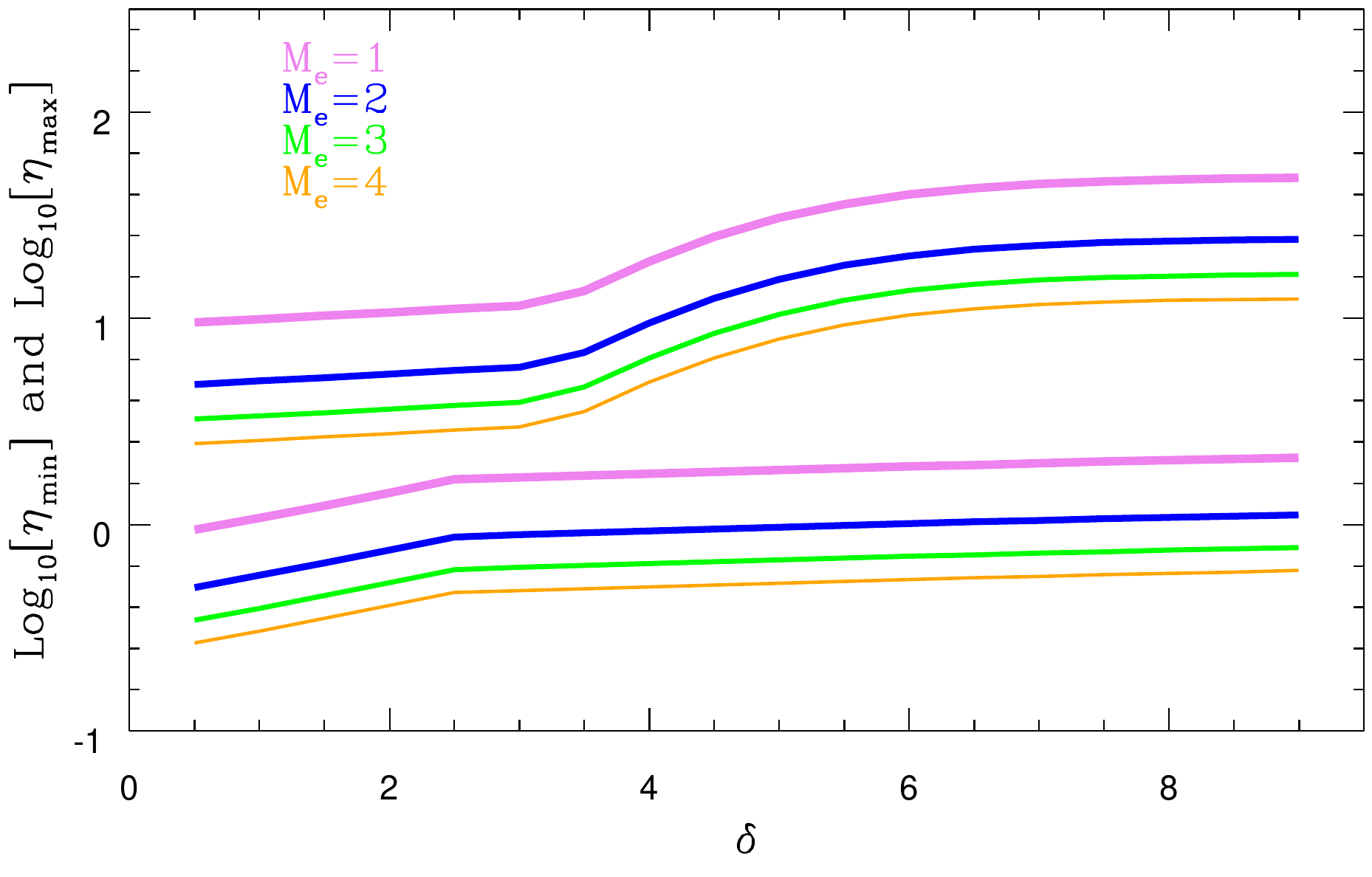}
    \caption{Values of $\log_{10}[\eta]$ are plotted as functions of $\delta$. Eighteen different values of $\delta$ were considered from $0.5$ to $9.0$. For each value of $\delta$, we consider four different  values of $M_\mathrm{e}$, the envelope mass. Each value of $M_\mathrm{e}$ is represented by 2 curves of a distinct color. For each pair, we computed $\eta$ for every WD-MS post-CE binary, and recorded the maximum and minimum values, $\eta_\mathrm{max}$ and $\eta_\mathrm{min}$, respectively. Plotted are $\eta_\mathrm{max}$ and $\eta_\mathrm{min}$ for each value of $\delta$.}
    \label{fig:delta}
\end{figure}

To explore the influence of $\delta$, we consider post-CE binaries containing a MS star in close orbit with a WD. In this case, our initial list, drawn from the post-CE catalog, consisted of 112 binaries. First we hold the envelope mass constant and explore the variation in $\eta$ among the post-CE binaries. For the purpose of computing $R_1(0)$, which is only weakly dependent on the initial donor mass, we use $R_1(0)(M_\mathrm{ZAMS}, M_\mathrm{c})$ from \citet{1987ApJ...319..180J}, where we set $M_\mathrm{ZAMS}$ to the sum of the core mass and the envelope mass. While the true value of $M_\mathrm{ZAMS}$ could  be larger if winds were active prior to RL filling, the trends we identify are not altered for different choices. In this set of calculations we vary the envelope mass between $1\, \Msun$ and $4\, \Msun$. 
We varied the value of $\delta$ from $0.5$ to $9$ in increments of $0.5$. For each value of $\delta$ we computed $\eta$ for each post-CE binary. We kept track of the minimum value of $\eta$, $\eta_\mathrm{min}$ and its maximum value $\eta_\mathrm{max}$. The results are shown in Figure \ref{fig:delta}. Note that for larger values of $\delta$, the factor $\mathcal{F}(\frac{M_\mathrm{c}}{M_\mathrm{2}},\delta)$ becomes smaller, so that the value of $\eta$ must be larger. Although we do not know exactly what values of $\eta$ should be assumed, we have noted that we expect it to be larger than roughly $0.1$ and smaller than $10$. Figure \ref{fig:delta} indicates that $\eta_\mathrm{max}$ begins to undergo a sharp increase at values of $\delta$ larger than $3$. The values of $\eta_\mathrm{min}$ decrease for lower values of $\delta$. Thus, suggests that we should use $\delta \approx 3$. We will we employ $\delta = 3$ throughout the remainder of the calculations described in this paper. We note, however, that the choice is not unique.

\subsection{The Value of \texorpdfstring{$\eta$}{eta}}
\label{sec:valueofeta} 

The value of $\eta$ depends on the mass ratios, $q=M_\mathrm{c}/M_\mathrm{2}$ and $q_\mathrm{ec}=M_\mathrm{e}/M_\mathrm{c}$.  Because the exponent contains the factor $\eta \times \frac{M_\mathrm{e}}{M_\mathrm{t}}= \eta\times q_\mathrm{ec} q/(1+q)$, where $M_\mathrm{t}=M_\mathrm{c}+M_\mathrm{2}$, we can see that, for a given post-CE binary with an unknown value of $M_\mathrm{e}$, larger values of $M_\mathrm{e}$ may be associated with smaller values of $\eta.$ This is simply a statement that, to produce a given final state, the specific angular momentum imparted to a portion of the envelop can be smaller when the total envelop mass is larger. The dependence of the factors outside the exponent means that the relationship between $\eta$ and $M_\mathrm{e}$ is somewhat more complicated, as we discuss in the context of different types of post-CE binaries.

We use observed post-CE binaries to estimate values of $\eta.$  The largest group of post-CE binaries for which we have reliable information is the set of 187 WD-MS binaries. In \S 4.4.1 we focus on them, and show similar figures for WD-WD and WD-NS systems in Appendix B.

\subsubsection{White-Dwarf/Main-Sequence Binaries}
\label{sec:WDMS}

\begin{figure*}
\begin{center}
    \includegraphics[width=\textwidth, clip, trim=0px 35px 0px 100px]{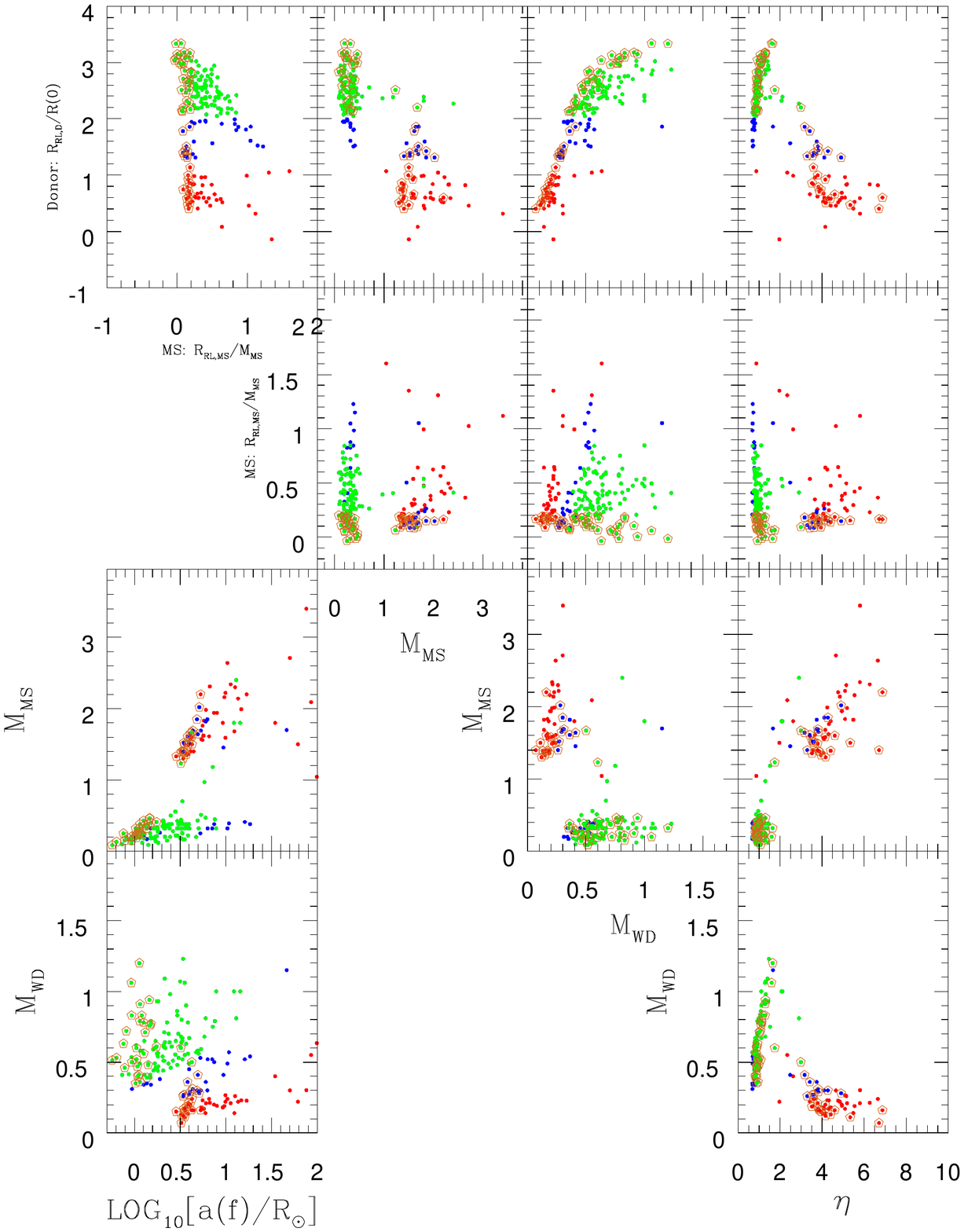}
    \caption{Relationships between the parameters of WD+MS post-CE binaries in the observational data catalogue from \citet{knd+21}. Green points correspond to binaries that have experienced significant shrinkage during a prior CE phase. Blue points have experienced more modest shrinkage, and red points have had little or no shrinkage. The quantities plotted are the WD and MS masses (in solar masses), and the logarithm to the base 10 of the final orbital separation (in solar radii).
    Also shown are $\log_{10}[\mathcal{R}_\mathrm{d}]$ (see equation \eqref{eq:rd}), a measure of shrinkage during the CE, and $\log_{10}[\mathcal{R}_\mathrm{MS}]$ (see equation \eqref{eq:rms}), a measure of whether the binary is close to filling its RL today. The values of $\eta$ have been estimated. Please see the text for more detailed explanations of the variables.}
    \label{fig:ce_ms}
\end{center}
\end{figure*}

We started with 187 WD-MS-star post-CE binaries (see Table~\ref{tab:obsdata}). To gain some insight into how these binaries were affected by the CE, we considered their possible initial states, i.e., the state before the binary shrank. For binaries in which a WD is the remnant of the RL filling companion, the initial radius of the donor, $R_1(0)$ must have been larger than its MS radius.

The binary's orbital separation at that time was $a(0)=R_1(0)/f(M_\mathrm{1}/M_\mathrm{2})$. We assume that the value of $M_\mathrm{2}$ does not change during the CE. $M_\mathrm{1}$, the mass of the donor, is the sum of the core mass, which is the mass of today's WD, and the envelope mass.  

By estimating the value of $M_\mathrm{e}$, we can estimate the value of $a(0)$. To do this, we have simply used the mean value of the possible range of envelope masses
\begin{equation}
    M_\mathrm{e}=0.5\, \left(M_\mathrm{e}^\mathrm{min} + M_\mathrm{e}^\mathrm{max}\right) ,
\end{equation}
where $M_\mathrm{e}^\mathrm{min}$ and $M_\mathrm{e}^\mathrm{max}$ are defined at the conclusion of \S~\ref{sec:stellarevolution}. Although these are not likely to be the true values of the envelope masses, by making the choice in the same way for all post-CE binaries we can make comparisons and examine trends. In Figure~\ref{fig:ce_ms}, the vertical axis in the top row of panels plots a variable $\mathcal{R}_\mathrm{d}$. Note that some of the red points appear to correspond to cases in which the radius \textsl{increased}. This could be due to uncertainties in the measured system parameters or in the applicability of our assumptions. Nevertheless, we cannot use the systems shown in red as good examples of post-CE binaries. Systems shown in blue have values of the ratio between 20 and 100. Some, possibly all of these, may have passed through a CE phase. Points in green have experienced shrinkage by a factor of 100 or more and are therefore the strongest candidates for post-CE systems.

In addition to these three colors we see that some points are surrounded by a golden pentagon. In these systems, the present-day separation is close enough that the MS star is filling or nearly filling its RL. While it is possible that the CE ended at the stage in which mass transfer from the MS star could start, this would appear to be a coincidence. More likely, the orbital separation shrank during a possibly long time interval after the CE had ceased. Thus, the systems represented by points surrounded by golden pentagons may not represent true post-CE states, but may instead be states that evolved after the end of the CE. Let $R_\mathrm{RL}^\mathrm{MS}$ represent the RL radius of the companion star with $M_\mathrm{2}$,
\begin{equation}
    R_\mathrm{RL}^\mathrm{MS} = \frac{a(f)}{f(M_\mathrm{2}/M_\mathrm{c})} .
\end{equation}
We use $M_\mathrm{MS}=M_\mathrm{2}$, the value of the MS-star's mass, as a proxy for the star's radius. Small values of the ratio
\begin{equation}
\label{eq:rms}
\mathcal{R}_\mathrm{MS}\equiv \left(\frac{R_\mathrm{RL}^\mathrm{MS}}{M_\mathrm{MS}}\right) {\rm ~~~\left[\frac{R_\odot}{M_\odot}\right]}
\end{equation}
are not favored for pure post-CE states. Thus, while we explore the possible evolution of all systems in the sample given in Table~\ref{tab:obsdata}, we view as pure CE survivors only the those green (and possibly blue) points not enclosed by pentagons, which correspond to systems
having $\mathcal{R}_\mathrm{MS}>1.5$.

We examine the trends exhibited by the distribution of systems in the panels of Figure~\ref{fig:ce_ms}, and what they tell us about the past and future evolution of these post-CE binaries. The vertical axis of the top set of panels shows $\mathcal{R}_\mathrm{d}$, the fractional shrinkage during the CE phase. Because of their definitions, the red, blue, and green points are in the bottom, middle, and top regions of the plot, respectively, corresponding to systems exhibiting the least (bottom) to most (top) shrinkage. The horizontal axis of the top left panel shows the ratio $\mathcal{R}_\mathrm{MS}$. Although there is a clear vertical grouping of systems that may be undergoing (or be close to undergoing) mass transfer, there are also systems (for each of the three colors) that are not currently interacting. $\mathcal{R}_\mathrm{MS}$ is also plotted along the vertical axis of the second row of panels. The left-most panel in the second row shows that there is a clear separation between binaries with MS stars having masses above $1\, \Msun$ and those with masses below roughly $0.6\, \Msun$. The green systems tend to house those MS stars of lower mass. In the middle panel of the second row we see that there is also a distinction between systems with higher-mass and lower-mass WDs, although these groups are not separated by a gap. Clear CE survivors (points in green) tend to have higher WD mass; note that some of this effect may be associated with their initially wider orbits.

This distinction is even clearer in the third-row plot of $M_\mathrm{MS}$ versus $M_\mathrm{WD}$. The clear CE survivors (green) tend to have higher-mass WDs and lower-mass companions. Their values of $q$ therefore are generally larger than for the systems that have experienced less shrinkage. Note however, that although the existence of clusters of points separated by mass is distinctive, it is not universal. There are some points spanning the gap.  By studying this panel, and the panels in which the component masses are plotted against the final (present-day) orbital separation, we can see that the systems lying between clusters of points may have important roles in the future evolution of binary populations. This is because they have WDs massive enough to be pushed to the Chandrasekhar mass by their companions, which could potentially each shed several tenths of a solar mass. We cannot say for certain that any of the WDs in this sparse set of systems will undergo accretion-induced collapse or SNe~Ia, but they are candidates for binaries that can do so via accretion. We must therefore ensure that any CE formalism is capable of predicting these systems and their numbers relative to other WD-containing binaries. 

Analyzing the binaries in the clusters that appear in these plots, we see that systems in red tend to have stars between $0.5\, \Msun$ and $3.5\, \Msun$ in orbit with helium WDs. Some of the orbits are close enough that mass transfer can start soon if it has not already started. Because the mass of the future donor (today’s MS star) is generally much higher than the mass of the helium WD, we expect that when the donor fills its RL, there will be a CE. Depending on the state of the donor star when that occurs, the result could be a system with either two helium WDs or a carbon oxygen and helium WD, which eventually merge subsequently \citep{dan14}. A similar fate could await some of the red-point systems with larger present day orbital separations. Systems in green have high mass WDs but their companion MS stars are of such low mass that they do not have much mass to donate. We would not expect the stars to experience a second CE phase, because the potential donor star is less massive than the mass of the potential accretor. These stars will likely become cataclysmic variables. Some of them may already be cataclysmic variables.

We can also study the post-CE groups with the goal of learning about the CE phase that produced them.  The systems in red were produced when a star with a helium core filled its RL. The shrinkage may have been minimal because the mass of the donor star was less than that of the companion, today’s MS star. Or it could have been the case that the small mass ratio, $M_\mathrm{1}/M_\mathrm{2}$, was associated with conditions that drive out the CE efficiently. Or else, the relatively dense envelope associated with a low-mass core in a donor of modest mass may be associated with conditions that are more easily driven out. The points in green, which experienced the most shrinkage had more evolved cores. The giant donor therefore evolved from stars that were initially more massive and were likely to be larger and less dense. The mass ratio was also larger. 

\subsubsection{Possible Values of $\eta$}

The fourth column of Figure~\ref{fig:ce_ms} plots as a function of $\eta,$ each of the quantities considered along the vertical axes described above $(\mathcal{R}_\mathrm{d}, \mathcal{R}_\mathrm{MS}, M_\mathrm{MS}, M_\mathrm{WD})$. To compute the value of $\eta$ for an individual system, we needed to select a value for $M_\mathrm{e}$. In these panels we used $M_\mathrm{e}=0.5\, (M_\mathrm{e}^\mathrm{min}+M_\mathrm{e}^\mathrm{max})$. In each panel, we see that the separation between the clusters in other panels is exhibited here as well. The certain CE survivors (green) have usually smaller values of $\eta$ than the systems in red that appear to have shrunk by smaller factors. Most of the points in green are smoothly spread along a curve whose smallest $\eta$ value is approximately $0.5$ and whose largest $\eta$ value is about $3$. Points in red have higher values, ranging up to about $7$.

We have carried out a parallel analysis of post-CE binaries that consist of NS-WD pairs, and also post-CE end states that consist of WD-WD pairs. The graphs are shown in the Appendix~\ref{sec:WDCO}. Because, as Table~\ref{tab:obsdata} shows, the numbers of systems in these categories are relatively small, detailed conclusions cannot be drawn. We note, however, that the values of $\eta$ are similar.

\subsubsection{Constant Values of $\eta$}
\label{sec:consteta}

If we wanted our formalism to apply to only the clearest examples of CE end states that form a relatively isolated clump, we could choose values of $\eta$ between $0.5$ and $2$. Under the assumption that trends discerned by using our chosen envelope masses are robust, this would leave out the CE end states that, as described above, may be related to accretion-driven SNe~Ia or accretion-induced collapse. It would also leave out those binaries which may be genuine CE end states, but which did not shrink as much. Indeed, the trends are likely to be robust, because, while different choices of envelope mass would move some of the points, the effect would be modest, since a factor of $n$ change in $\eta$ generally requires a multiplicative change of similar magnitude in $M_\mathrm{e}$.

\begin{figure}
\begin{center}
    \includegraphics[width=\columnwidth, clip, trim=20px 40px 30px 60px]{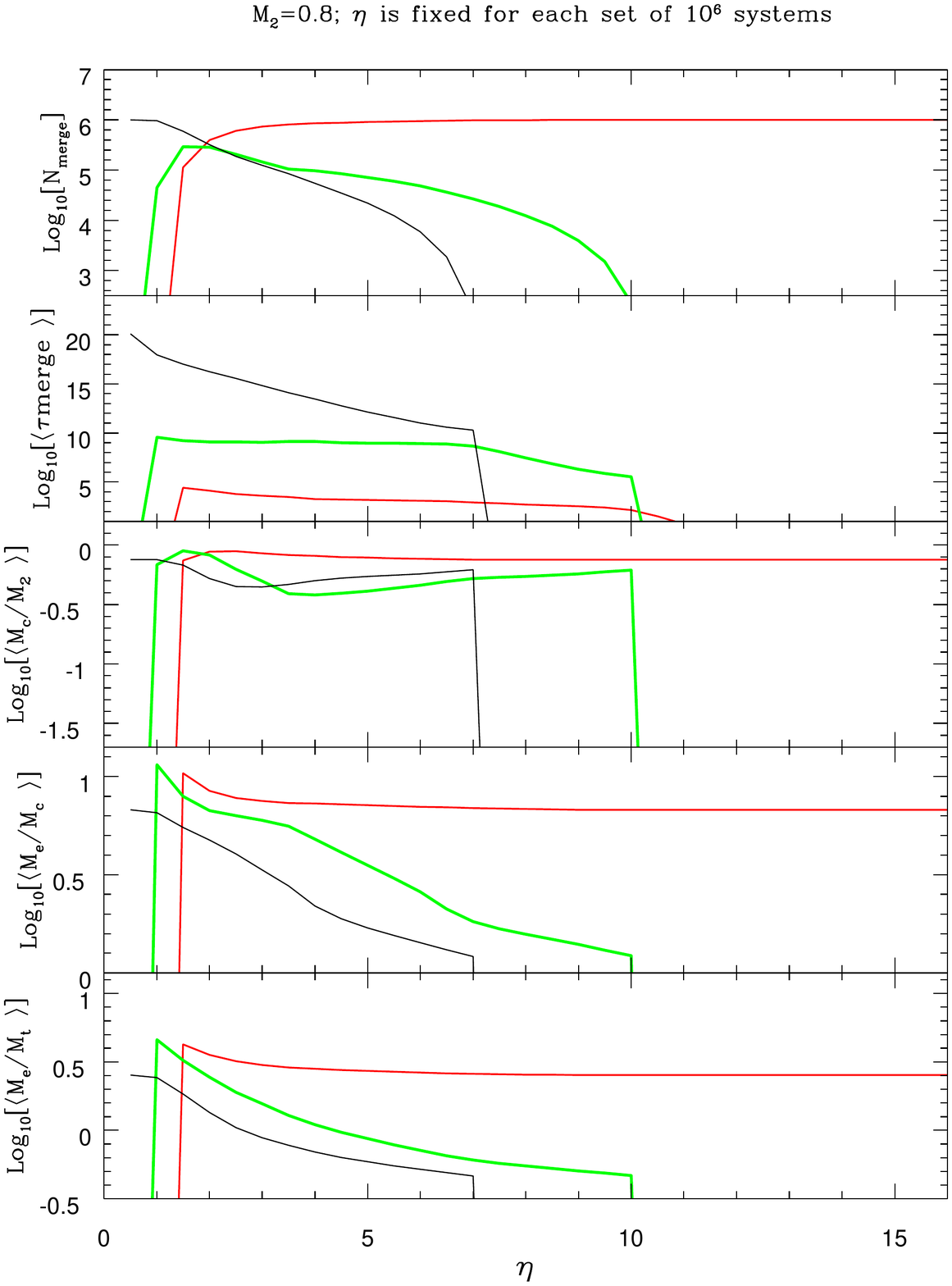}
    \caption{Curves in red correspond to systems that merge within $\unit{2\times 10^5}{\years}$ in our simulation; systems represented in green which merge within a Hubble time; in black are systems that require times longer than the Hubble time to merge. The value of $M_\mathrm{2}$ is set to be $\unit{0.8}{\Msun}$. The value of $\eta$ was held fixed for a simulation of $10^6$ CE evolutions, as described in the text.}
    \label{fig:stat_m2-0_8}
\end{center}
\end{figure}

\begin{figure}
\begin{center}
    \includegraphics[width=\columnwidth, clip, trim=20px 40px 30px 60px]{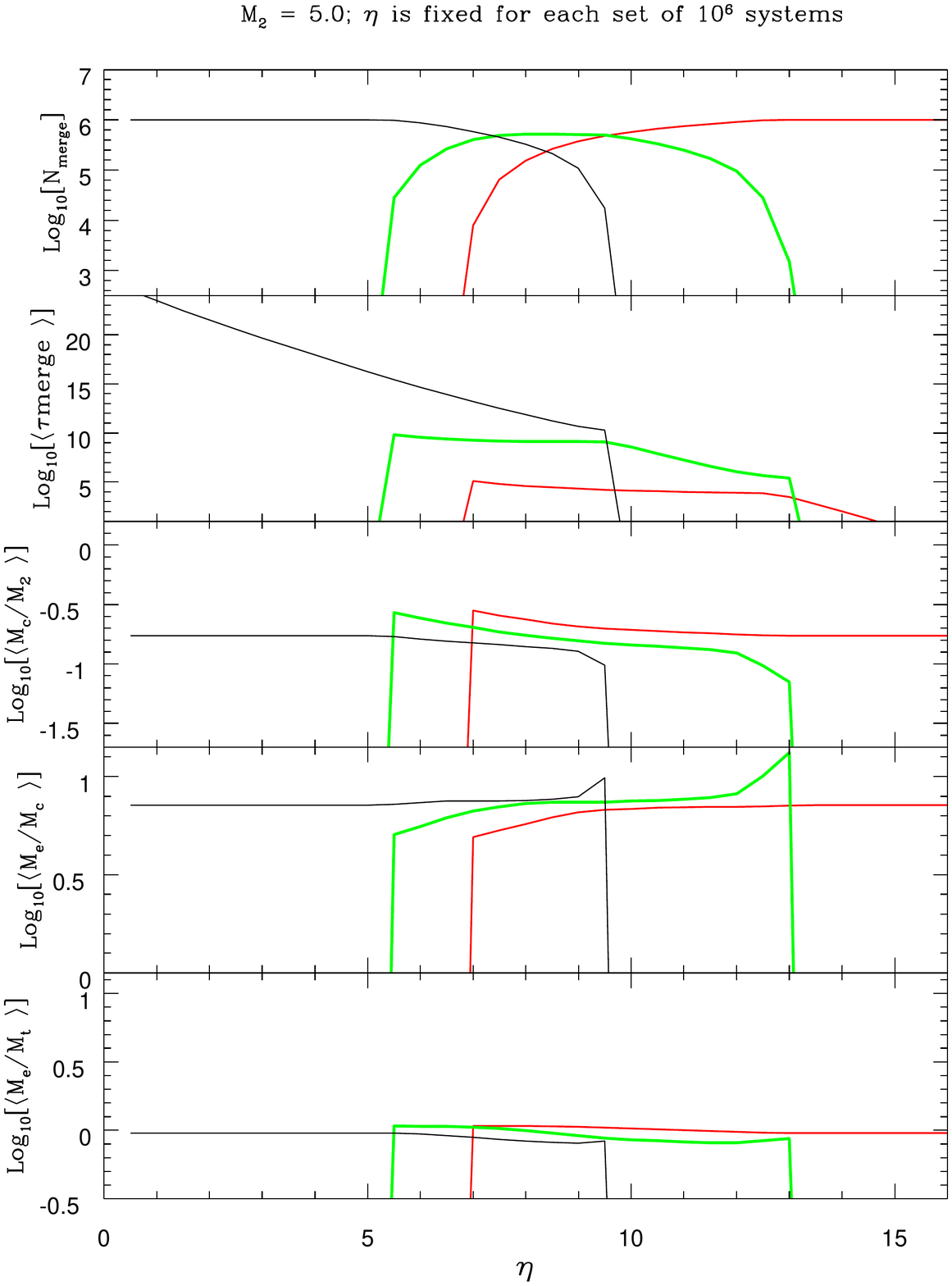}
    \caption{Similar to Figure~\ref{fig:stat_m2-0_8} but with $M_\mathrm{2}=\unit{5}{\Msun}$.}
    \label{fig:stat_m2-5}
\end{center}
\end{figure}

We have therefore conducted a set of simulations that shows how different values of $\eta$ influence CE end states. Starting with a fixed value of the companion's mass, $M_\mathrm{2}$, we consider a population of $\sim 10^6$ binaries that will undergo a CE phase. We take $\eta$ to have a constant value and conduct the same simulation
for 33 different values of $\eta$. This allows us to assess the effects of selecting different values of $\eta.$ In order to quantify the result, 
we focus on estimating the rates of gravitational mergers post CE. At the time the envelope is released, the mass of the donor star, $M_\mathrm{1}$, is the sum $M_\mathrm{c}+M_\mathrm{e}$. The companion has mass $M_\mathrm{2}$, which is assumed to remain constant during the CE.
We begin the calculation by selecting a specific value for the companion's mass $M_\mathrm{2}$ to be $\unit{0.8}{\Msun}$ ($\unit{5}{\Msun}$) to create the plots in Figure~\ref{fig:stat_m2-0_8} (Figure~\ref{fig:stat_m2-5}).
The donor star must be more massive than $M_\mathrm{2}$. We chose a value between $M_\mathrm{2}$ and $\unit{8.4}{\Msun}$, using a uniform distribution. Considering that, at the time of RL filling the donor may have lost some mass, we define $M_\mathrm{WD}^\mathrm{max}$ to be the largest possible WD mass, and select $M_\mathrm{c}$ to have a value between $\unit{0.1}{\Msun}$ and $M_\mathrm{WD}^\mathrm{max}$, using a uniform distribution. 

The donor's mass and core mass allow us to estimate the radius of the donor (hence the RL radius) at the time of RL filling. The RL radius and the masses of the donor and companion provide an estimate of the orbital separation at the time the CE started. The envelope mass was the difference between the donor's mass and the mass of its core. Considering only those binaries in which the envelope mass as computed here lies within the range defined by the minimum and maximum expected values of $M_\mathrm{e}$, we had all of the information needed to compute $a(f)$ and $\tau_{\mathrm{merge}}$.

For each of 33 values of $\eta$ in the range 0.5 to 16, we mapped the evolution of $10^6$ binaries through the CE phase. We counted the numbers of systems that merge within $\unit{2\times 10^5}{\years}$ (red), the number that merge within a Hubble time (within $\unit{1.37\times 10^{10}}{\years}$, green), and the number that take longer than a Hubble time to merge (black). The first category was created to explicitly consider the proportion of systems that may merge before the CE has a chance to disperse, since these may exhibit distinctive features. Those that merge within this short interval or else within a Hubble time, may be used to compute the rates of potentially observable events. 

The top panel in Figures~\ref{fig:stat_m2-0_8} and \ref{fig:stat_m2-5} shows the logarithm to the base 10 of the number of mergers. Each subsequent panel shows the logarithm to the base 10 of the average value of one quantity: the time to merger (in years), the mass ratio $q=M_\mathrm{c}/M_\mathrm{2}$, the mass ratio $M_\mathrm{e}/M_\mathrm{c}$, the mass ratio $M_\mathrm{e}/(M_\mathrm{c} + M_\mathrm{2})$. To explore the effect of increasing $\eta$, we can start at the left of each panel and move toward the right. 

Clearly, higher values of $\eta$ produce more frequent and earlier mergers. The average mass ratio of the post CE binary is not monotonic. Usually, the core of the donor is always less massive than the companion star. 
In general, the envelope masses are larger than the core masses at the onset of our simulated CE. This even holds when comparing the envelope to the total post CE binary mass instead of the donor's core mass. As we fixed the companion mass, the two lower panels of Figure~\ref{fig:stat_m2-0_8} show the same trend of having lower envelope masses for larger values of $\eta$. 
Thus, the cutoff at lower $\eta$ values is caused by the choice of the maximally allowed envelope mass. Vice versa, the cuts at higher values of $\eta$ are related to the minimum envelope mass, which is often given by the requirement that the donor mass should be larger than the companion mass. By comparing Figures~\ref{fig:stat_m2-0_8}, for which $M_\mathrm{2}=\unit{0.8}{\Msun}$, and \ref{fig:stat_m2-5}, for which $M_\mathrm{2}=\unit{5.0}{\Msun}$, find that larger values of $\eta$  are required to achieve mergers when the companion mass is larger. For example, there is a limited range for $\eta$ where the three groups according to the merger time have comparable values, and that this value increases as $M_\mathrm{2}$ increases.

\subsubsection{Challenges of employing a fixed value of $\eta$}

When conducting population synthesis calculations, we start with many individual binaries and follow the evolution of each. For those binaries that, at some time during their evolution, start a CE, we must decide what the value of $a(f)$ will be. For parametric formalisms, including the $\alpha$ formalism for example, the simulation may be run multiple times, with different CE parameter values selected for each simulation. Thus, for example, a constant value of $\alpha \, \lambda$ may be used in each simulation, and the results of different simulations are compared to test the effects of employing different values of the CE parameter(s).

If we consider a set of simulations using a single value of $\alpha \, \lambda$, we know that the value employed is likely to be a good approximation of the true value for a subset of the binaries. The computed CE final states of these binaries would be similar to the final states that would be produced in nature.  But for most other binaries, the computed and true final states would be different. Nevertheless, a set of simulations which employs a different value for the product $\alpha \, \lambda$
would yield correct results for a different set of binaries. 
If we understood the CE phase, we would know which values of $\alpha\, \lambda$ are appropriate for any given binary.
Without this knowledge, when we use different parameter values for different simulations, the meaning of the comparisons 
between simulations is difficult to untangle.
The same is naturally true if we choose to use a single value of $\eta$ in a specific simulation, and vary the values among simulations.\footnote{The same could in principle be said about employing a fixed value of $\delta$, but, as shown above, we
have selected a regime in which the results should be robust with respect to modest changes in $\delta.$}

The positive feature associated with using a fixed value of the CE parameters is that it simplifies the input to the simulations. If, however, we had a simple functional form to express the CE parameters in terms of the binary's physical characteristics, we could maintain
the simplicity while achieving a more readily interpreted set of population synthesis results. A formula-based approach has been taken in other
formalisms \citep{demarco11}. 
In the rest of this paper we explore a trend that allows us to derive and use a functional form for $\eta.$

\section{Functional Form for $\eta$}
\label{sec:functioneta}
\begin{figure}
    \centering
    \includegraphics[width=\columnwidth, clip, trim=30px 40px 30px 100px]{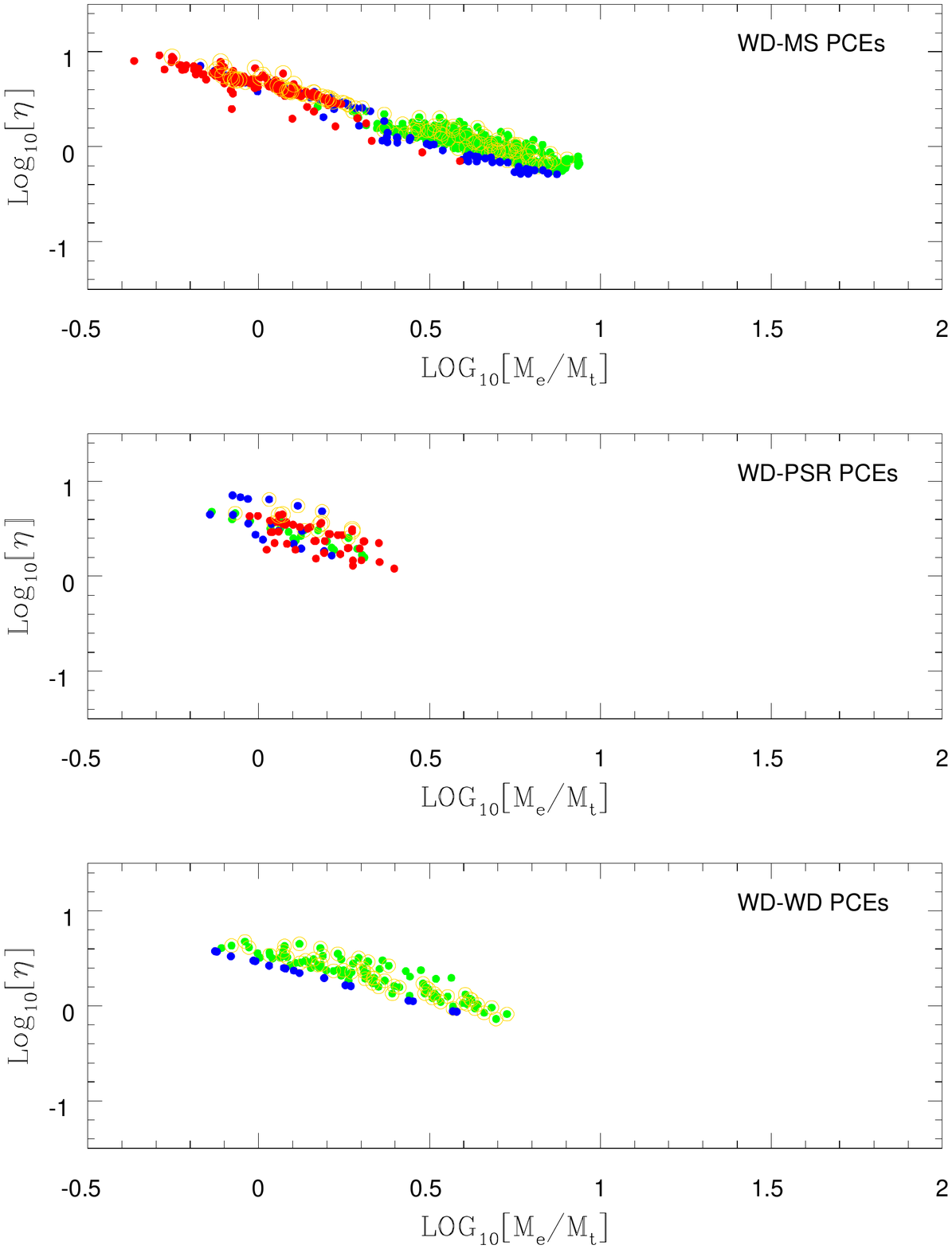}
    \caption{The dependence of $\eta$ on $M_\mathrm{e}/M_\mathrm{t} = q_\mathrm{ec} q/(1+q)$. Note that $\eta$ is a function of only $q$ and $q_\mathrm{ec}$. The different panels are for different types of post-CE systems (PCEs) from top to bottom: WD-MS, WD-NS, double WD. The colors and symbols are the same as in Figure~\ref{fig:ce_ms}. Here each observed system is represented by three data points using the first, second, and third quartile between $M_\mathrm{e}^\mathrm{min}$ and $M_\mathrm{e}^\mathrm{max}$ for the envelope mass.}
    \label{fig:eta_2}
\end{figure}

Figure~\ref{fig:eta_2} shows $\log_{10}[\eta]$ plotted against $\log_{10}[M_\mathrm{e}/M_\mathrm{t}]$. 
Note that 
\begin{equation}
    \frac{M_{{\mathrm e}}}{M_{{\mathrm t}}} = \frac{q_\mathrm{ec} q}{1+q} ,
\end{equation}
where $M_{\mathrm t}=M_{\mathrm c}+M_{\mathrm 2}$.
Thus, this graph shows $\eta$ as a function of a variable involving only $q$ and $q_\mathrm{ec}$.
In the top panel, the points correspond to WD-MS post-CE binaries. The points exhibit a linear trend, which is also consistent with the more sparsely populated middle and bottom panels,
 showing WD-NS and WD-WD systems, respectively. 
 
 Each system  here is represented by three data points corresponding to three choices of the envelope mass.
To select the values for $M_\mathrm{e}$, we computed the maximum and minimum envelope mass (as in the previous subsection) for each donor star. We then divided this interval into four equal pieces, and selected three values of  $M_\mathrm{e}$:
the middle value and the ones just above and below the middle value. Thus, we excluded only the minimum and maximum possible values of the envelope mass. The physical system may have had an envelope mass close to one of these values, but not necessarily. If we knew the actual value of the envelope mass for each system, we would be able to re-plot the figure and determine the most appropriate functional form. Without the ability to do that, we can use the points in Figure~\ref{fig:eta_2} as a guide to the trend that expresses the functional form in an approximate way as follows: 
\begin{equation}
    \label{eq:fit}
    \log_{10}[\eta] = -A\, \log_{10}[M_\mathrm{e}/M_\mathrm{t}] + B .
\end{equation}
In the case of $A=0$, $\eta$ becomes constant independent of binary parameters.
To derive the fit parameters ($A$ and $B$) we employ a straight line model. Given the data at hand, we cannot justify a more complicated model. 
Whether or not a straight-line fit is the optimal choice, the fact some functional form links $\eta$ to a set of binary-system parameters is important. It allows us to eliminate $\eta$ from the equations for $a(0)/a(f)$. For each set of post-CE binaries, the ratio of initial to final separations can be expressed as a function of $q$ and $q_\mathrm{ec}$. BPS calculations can compute post-CE separation analytically.
This simplification  makes the formalism relatively straightforward to implement, an important consideration for BPS calculations.

We now have
\begin{equation}
    \frac{a(f)}{a(0)} = \left[\frac{\left( 1+q \right) \left(1+q_\mathrm{ec} \right)^2}{1+q\, \left( 1+q_\mathrm{ec} \right)}\right]\, e^{\left[ -2 \mathcal{F}(q, \delta) \, \left(10^{B+\Delta} \times \left(\frac{q_\mathrm{ec} q}{1+q}\right)^{1-A}\right)\right]} .
\end{equation}
We see that only the values of $q$ and $q_\mathrm{ec}$ are needed in order to compute $a(f)/a(0)$. There is no explicit dependence on $\eta$, which has been replaced by it's functional form. 
The values of $A$ and $B$ are determined by the type of binary we are considering. 
We see that larger values of $B$ produce smaller final orbits. Variations in $A$ can produce complex behaviors.
For $A=1,$
the exponent has no dependence on $M_\mathrm{e}$. For $A<1$, orbital shrinkage is facilitated as $q_\mathrm {ec}$ and as $A$
decreases. For $A>1$ orbital shrinkage is moderated as $A$ increases and also as $q_\mathrm{ec}$ increases.
In the expression on the right-hand-side, we have 
added a term $\Delta$ to $B$. When conducting simulations, we can use this term to generate the spread associated with the residuals by selecting $\Delta$ from a Gaussian distribution with width $\sigma.$ 

Fits to the data in Figure~\ref{fig:eta_2} produce 
values for the parameters $A$ and $B$ that are given in Table~\ref{tab:fits}.  Although $\sigma$, also in the table and derived from the distribution of residuals,  provides a rough estimate of the uncertainties,
a deeper source of uncertainty is that we do not know which choice of the envelope mass is best.
In separate work we consider possible scenarios in which the envelope masses and parameters can be chosen in a mutually consistent way.

Table~\ref{tab:fits} shows the results for the combined data (``All'', top row), as well as the fits for six types of post-CE binary.
Note that, in this row, $A\approx 1$ and $B\approx 0.6$. 
These  values are close to those found in all cases with small to modest shrinkage (${\cal R}_{\mathrm d} < 20$; rows 3, 5, and 7 in Table~2). 

\begin{table}
    \caption{\label{tab:fits} Functional form of $\eta$: fits of Equation~\eqref{eq:fit} to plots in Figure~\ref{fig:eta_2}. The last column is the approximated standard deviation of the residuals.}
    \centering
    \small
    \begin{tabular}{cccc}
    \hline
        Type & Slope, $A$ & Intercept, $B$ & $1\, \sigma$ \\
        \hline
        All &   $0.952 \pm 0.011$ & $0.603 \pm 0.006$ & $0.08$ \\
        WD+MS (Shrink $> 20$) &   $0.892 \pm 0.021$ & $0.567 \pm 0.014$ & $0.08$ \\
        WD+MS (Shrink $< 20$) &   $1.027 \pm 0.047$ & $0.637 \pm 0.008$ & $0.08$ \\
        WD+NS (Shrink $> 20$) &   $1.285 \pm 0.072$ & $0.678 \pm 0.024$ & $0.08$ \\
        WD+NS (Shrink $< 20$) &   $1.051 \pm 0.120$ & $0.561 \pm 0.020$ & $0.11$ \\
        WD+WD (Shrink $> 20$) &   $0.780 \pm 0.046$ & $0.487 \pm 0.027$ & $0.065$ \\
        WD+WD (Shrink $< 20$) &   $0.969 \pm 0.130$ & $0.640 \pm 0.034$ & $0.03$ \\
        \hline
    \end{tabular}
\end{table}

\begin{figure*}
\begin{center}
    \includegraphics[width=\textwidth, clip, trim=0px 30px 0px 100px]{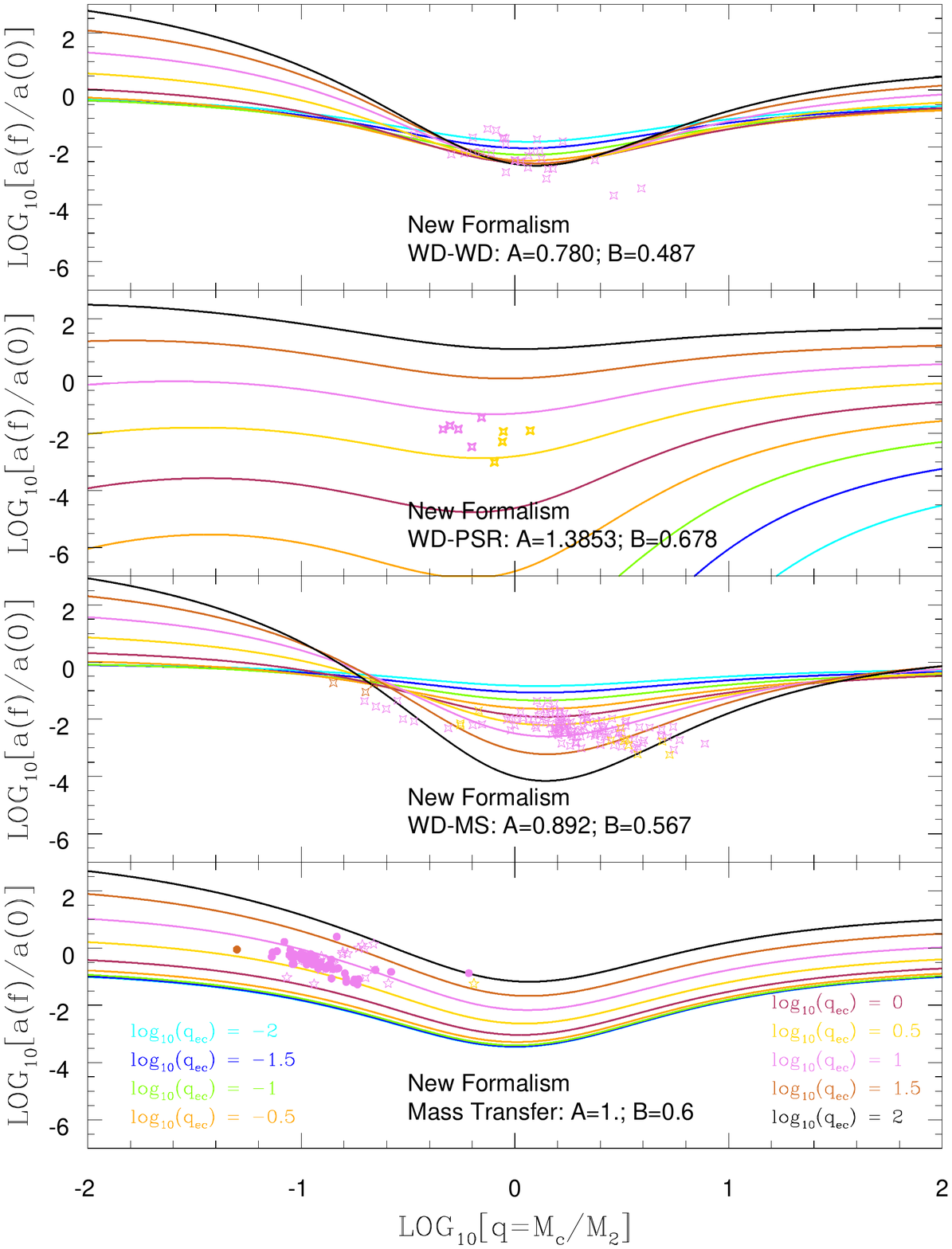}
    \caption{In the new formalism, the ratio $a(f)/a(0)$ is shown as a function of the final mass ratio $q=M_\mathrm{c}/M_\mathrm{2}$. The only other variable is $q_\mathrm{ec}=M_\mathrm{e}/M_\mathrm{c}$, which is sampled by the separate curves as indicated in the bottom panel. The points in each panel represent a particular set of the post-CE binaries; the top three panels are for double WD, WD-NS, and WD-MS, with a substantial shrinkage, cf. blue and green points in Figure~\ref{fig:ce_ms}. The bottom panel only contains binaries with $\mathcal{R}_\mathrm{d} < 20$, cf. red points in Figure~\ref{fig:ce_ms}.} 
    \label{fig:ce_shrink_eta}
\end{center}
\end{figure*}

Figure~\ref{fig:ce_shrink_eta} plots curves of $\log_{10}[a(f)/a(0)]$ versus $q$, for each of nine values of $q_\mathrm{ec}$ (color-coded), given in the bottom panel. We also show points superposed on each panel. These correspond to the post-CE binaries most likely to have actually experienced a significant CE episode. 

In the bottom panel we use $A=1$ and $B=0.6$. Table~\ref{tab:fits} shows that, whatever the composition of the binary, the systems that have had little shrinkage ($\mathcal{R}_\mathrm{d} < 20$) have values of $A$ and $B$ close to $1.0$ and $0.6$, respectively.  The curves in this panel are non-intersecting. Larger values of $q_\mathrm{ec}$, which must at least sometimes correspond to large $M_\mathrm{e}$, exhibit less shrinkage, and some even expand. The points superposed are the post-CE binaries that have ($\mathcal{R}_\mathrm{d} < 20$);  they include WD-MS, WD-NS, and WD-WD binaries, drawn from rows 3, 5, and 7 of Table~2. . 
This panel demonstrates that, because SCATTER's physical assumptions are closely tied to angular momentum conservation,
modest orbital shrinkage, and even orbital expansion, are covered by the SCATTER formalism. The fact that these types of post-CE candidates have positions encompassed by the curves in this panel may indicate that there is a relatively smooth
transition between some CE systems and binaries experiencing stable RL overflow. 

Row 2 in the table corresponds to WD-MS binaries with ${\cal R}_{{\mathrm d}} >20$. These are likely CE survivors.  Recall that WD-MS binaries are post-CE binaries with a WD in orbit with a main-sequence star. The pre-CE system to which the equation applies is a MS-giant binary. Rows 4 and 6 also show binaries with ${\cal R}_{{\mathrm d}} >20$, but for WD-NS and WD-WD binaries, respectively. The values of $A$ and $B$ are different for each type
of post-CE binary.

The second panel from the bottom in Figure~\ref{fig:ce_shrink_eta} panel shows WD-MS post-CE binaries with $\mathcal{R}_\mathrm{d} > 20$. This panel applies to the largest number of post-CE binaries in any single category, shown superposed in the analytic curves. We see that for $A<1$, there can be expansion of the CE binary for low values of $q$, and that the expansion is associated with high envelope mass. As $q$ increases, however, the curves cross and that the binaries which shrink most have large values of $M_\mathrm{e}$. Most of the data points fit well within the region where shrinkage is expected, with $a(f)/a(0)$ going down as low as $10^{-3}$. A few points are just outside the curves, and seem to exhibit even more shrinkage. Thus may indicate that the value of $A$ and/or $B$ found through the fit is too small, or it may simply reflect our uncertainty.

The top two panels correspond to binary types that were not well represented in our set of systems: WD-NS systems in the third panel from the bottom and WD-WD systems in the top panel. Given the small numbers of WD-WD and WD-NS systems, it is difficult to quantify the uncertainty. We have used $\Delta$, which in this paper we have taken simply to be $\sigma$, but as we point out in \S~\ref{sec:applications}, the true uncertainties are likely to be larger.

\section{Application}
\label{sec:applications}

\begin{figure}
\begin{center}
    \includegraphics[width=\columnwidth, clip, trim=30px 40px 30px 100px]{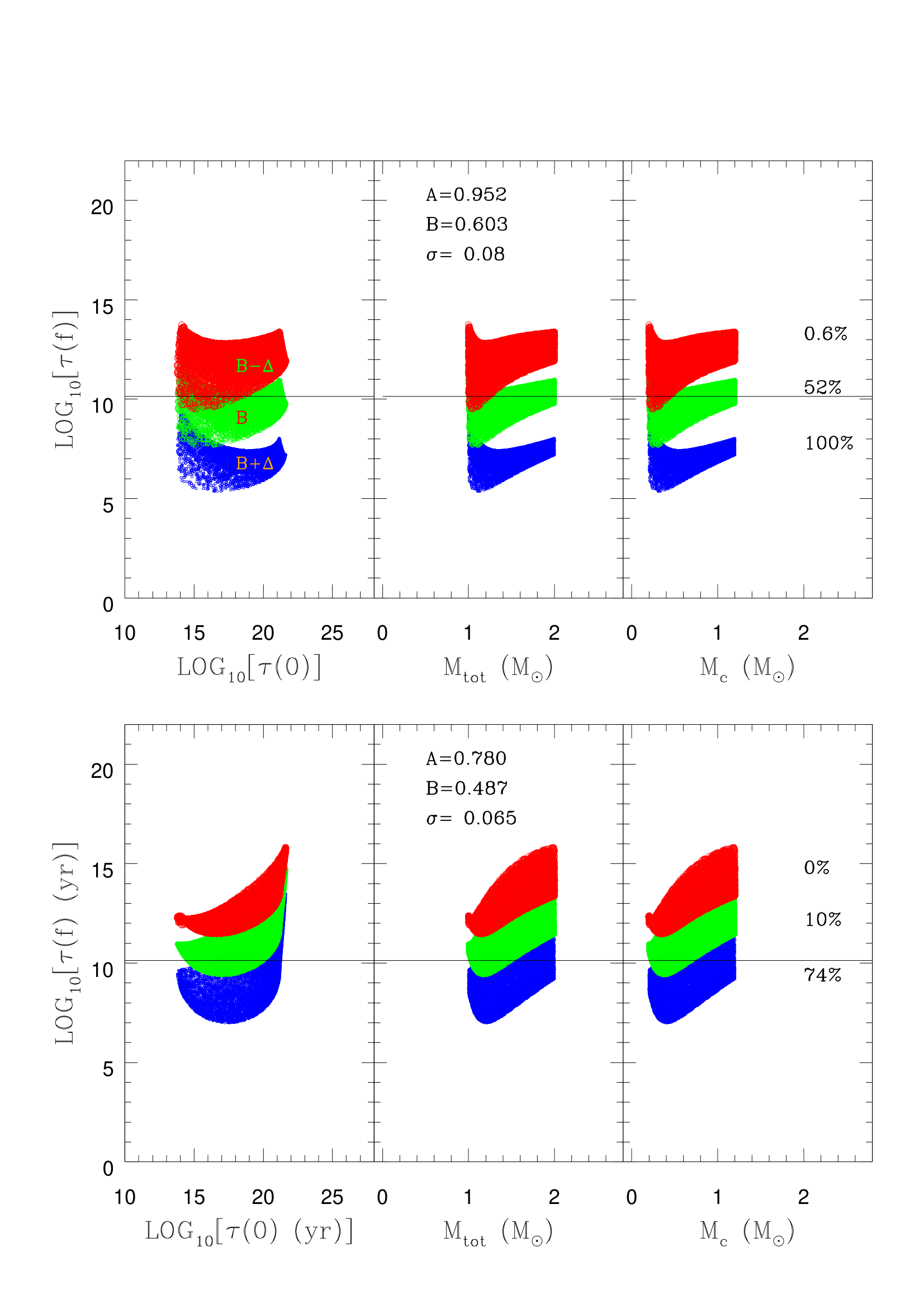}
    \caption{For WD-WD binaries, times to merger through gravitational radiation after the CE, plotted against the pre-CE time to merger, the total binary mass, and the mass of the donor's core. The parameter values are shown in the figure and described in the text. Green: the parameters used are $A$ and $B$ from the middle panel in the row. Blue: $B$ is replaced by $B+\Delta$ with $\Delta=1\, \sigma.$
    Red: $B$ is replaced by $B-\Delta$ with $\Delta=1\, \sigma.$}
    \label{fig:ce_population_low_mass}
\end{center}
\end{figure}

As discussed in the introduction, SNe~Ia progenitor models generally involve binaries which undergo one or two CE phases. 
In the case of two phases, the second CE typically leads to a
double-WD state. If the two WDs come close enough together that one of them
will fill its RL, there could be a merger. Several types of mergers
(e.g., both CO-CO mergers and CO-He mergers) may be able to produce
SNe~Ia. The time to merger after the CE phase can be estimated by using
equation~\eqref{eq:tmerge}. 

The goal of this paper is not to assess the likelihood of mergers within a realistic
stellar population, but instead to 
study how the SCATTER formalism works in producing close WDs, and to develop a better understanding of the uncertainties. To do this,
we generated binaries containing a single WD and a companion star that fills its RL at a point when mass transfer will be dynamically unstable. The  RL filling condition determines the pre-CE orbital separation. We started with a WD of $\unit{0.8}{\Msun}$, and then introduced a more massive companion ($\unit{0.8}{\Msun} < M_\mathrm{1} < \unit{8.4}{\Msun}$ with an evolved core ($M_\mathrm{c} > \unit{0.1}{\Msun}$). We chose the mass values from uniform distributions. 

Starting with the initial states described above, we used the SCATTER formalism to compute the post-CE separation between the donor's core, which itself becomes a WD, and the first-formed WD. Thus, the resulting system is 
comprised of two WDs in a binary whose orbital radius we have computed. We have enough information to compute the time to merger, assuming that the post-CE binary has a circular orbit. The results are shown in Figure~\ref{fig:ce_population_low_mass}. We do not make any assumptions about which pairs of WDs can produce SNe~Ia. (See \citealt{soker19} and \citealt{ruiter20} for a summary of possible channels.)

The primary goal of this exercise is to study the uncertainties, which are related to our uncertainties in the values of $A$ and $B$. In particular, because there were relatively few WD-WD systems among the post-CE binaries in our sample, the values of $A$ and $B$ are more uncertain than for WD-MS binaries. Additionally, when considering the observed WD-WD post-CE candidates, we did not know which WD had evolved first, and assumed that it was the more massive WD; this may not in fact have been the case for all  binaries in our sample. We therefore used two sets of values for the pair $(A, B)$, as shown in the panels of Figure~\ref{fig:ce_population_low_mass}. 

In the top panels we used values close to the average for the entire sample ($A=0.952, B= 0.603)$, with $\sigma=0.08$. In the bottom panels we use the values derived specifically for WD-WD binaries  ($A=0.780, B= 0.487)$, with $\sigma=0.065$.  Mergers taking place within a Hubble time are represented by systems lying below the black line in each panel. 
The parameters we derived based on WD-WD post-CE binaries produced fewer mergers within a Hubble time.

Even if we have a well-sampled group of binaries for the WD-WD model, there is still
an uncertainty based on the residuals to the fit. Thus, when we compute $A$ and $B$ for
a given system that will become a close double-WD binary, we must take into account the uncertainty $\sigma.$ In the context of a population synthesis simulation,  for a given system, $i$, we would take: $B=B+\kappa \times \sigma$, where we choose $\kappa$ from a Gaussian distribution of width $\sigma$.
Thus, even for a single binary in the simulation, a range of possible values of $a(f)$ are possible. This uncertainty is derived based on the actual post-CE binaries with which we started.  It reflects a genuine and expected uncertainty due to the fact that,
even if the variables we have used for two different binaries are identical, there will be some physical differences between the binaries that should produce a spread in the values of $a(f).$

To explore the differences, we consider, for each pair $(A, B)$, what happens if we 
use $B+\Delta$, where we have selected $\Delta=\sigma$ and also $B-\Delta$.
We find that larger effective values of $B$ always produce smaller final-state binaries, as the 
formula indicates they should.

We will show in the next section that the changes in separation predicted by the SCATTER formalism can be modest, compared with those predicted by other formalisms. It is therefore important to note that in each case, a significant number of mergers are expected within a Hubble time. 
In the upper (lower) panels $100\%$ ($74\%$) of all binaries for which $B$ is replaced by $B+\Delta$ 
merged in a Hubble time. We would expect roughly a third of all binaries in the simulations to have low values of $B$. In addition, some binaries ($52\%$ in the upper panels, $10\%$ in the lower panels) also merged. 

In the bottom, taken from the WD-WD post-CE binaries, the mergers tended to have a lower-mass WD (possibly a helium WD), merging with a WD of higher mass. In the upper panel, the masses of the merging WDs were on average higher. Both situations provide potential channels to SNe~Ia.


\section{Comparisons with other Formalisms}
\label{sec:comparison}

\subsection{The \texorpdfstring{$\alpha$}{Alpha} Formalism}
\label{sec:compalpha}

The so-called $\alpha$ formalism \citep{web84, dek90} is a traditional and more common way the deal with a CE. It is based on energy conservation. In its original form it assumed that orbital energy is the only source providing the energy needed to unbind the envelope material. The formalism contains two parameters:
$\alpha$ describes the efficiency of the energy conversion; $\lambda$  describes the strength of envelope binding. In the final equation, which describes the change of the orbital separation, those two parameters  appear only in a product, so that
some studies employ the product as the only free parameter.
\begin{equation}
    \label{eq:aral}
    \frac{a(0)}{a(f)} = \left(1+q_\mathrm{ec}\right)\, \left(1+\frac{2\,q_\mathrm{ec}\,q}{\alpha\,\lambda}\,\left[f(q+q_\mathrm{ec}\,q)\right]^{-1}\right) .
\end{equation}

\begin{figure}
\begin{center}
    \includegraphics[width=\columnwidth, clip, trim=30px 40px 30px 100px]{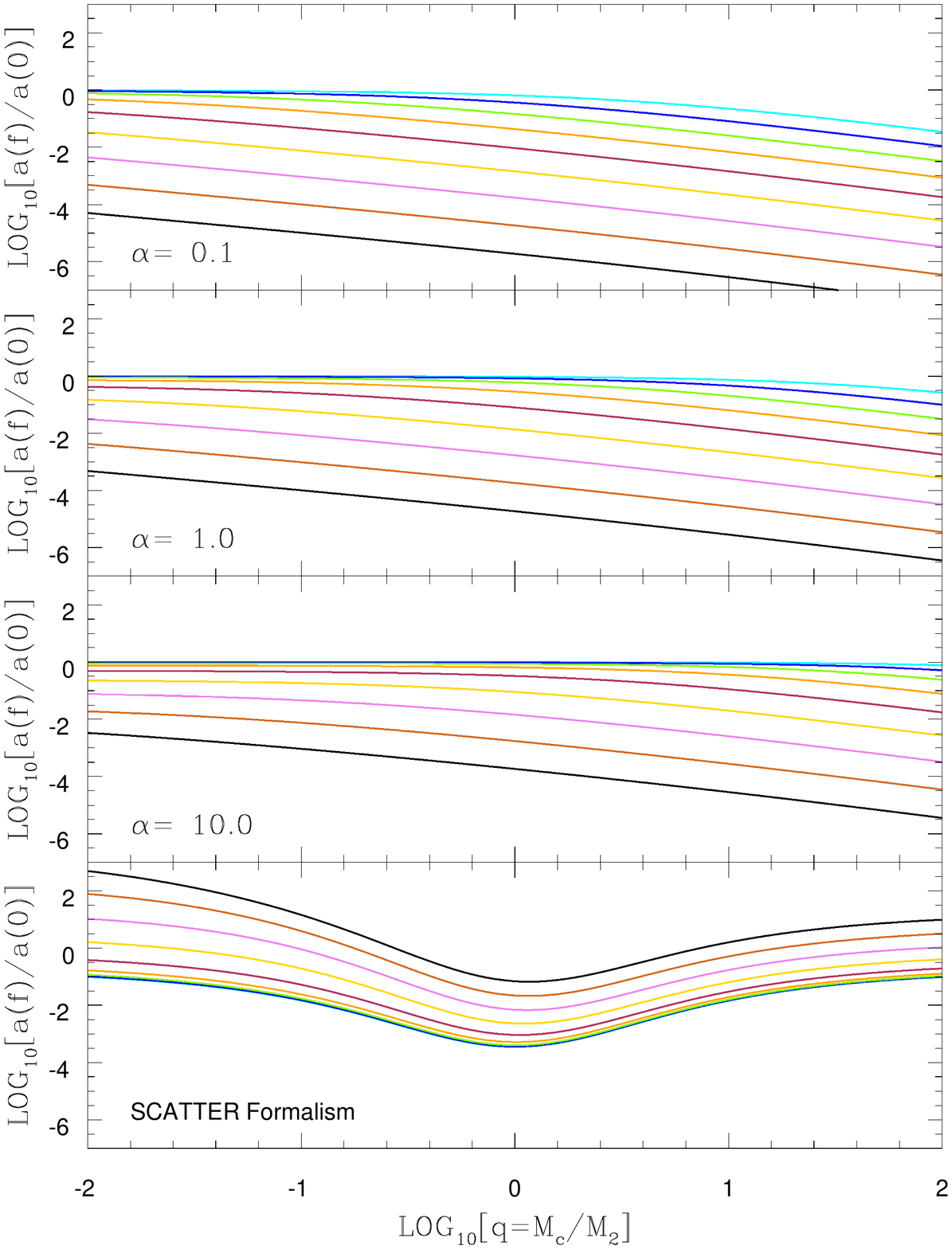}
    \caption{In both the $\alpha$ formalism (varying $\alpha \, \lambda$) and the SCATTER formalism for post-CE systems, the ratio $a(f)/a(0)$ is shown as a function of the final mass ratio $q=M_\mathrm{c}/M_\mathrm{2}$. The only other variable is $q_\mathrm{ec}=M_\mathrm{e}/M_\mathrm{c}$, which is sampled by the separate curves as shown in the bottom panel of Figure~\ref{fig:ce_shrink_eta}.}
    \label{fig:ce_shrink_alpha}
\end{center}
\end{figure}

In Figure~\ref{fig:ce_shrink_alpha} we compare the SCATTER formalism with the $\alpha$ formalism, using three values of the product $\alpha\, \lambda$ sometimes applied in simulations: $\alpha \, \lambda = 0.1, 1, 10$. There are several differences between the formalisms.

First, the SCATTER formalism (in common with the angular-momentum based $\gamma$ formalism discussed below), allows for orbital expansion, as well as for the shrinking of the orbit.
This is because orbital evolution during the CE phase depends on the amount of angular momentum carried by mass exiting the system. 

\begin{figure}
\begin{center}
    \includegraphics[width=\columnwidth, clip, trim=30px 40px 30px 100px]{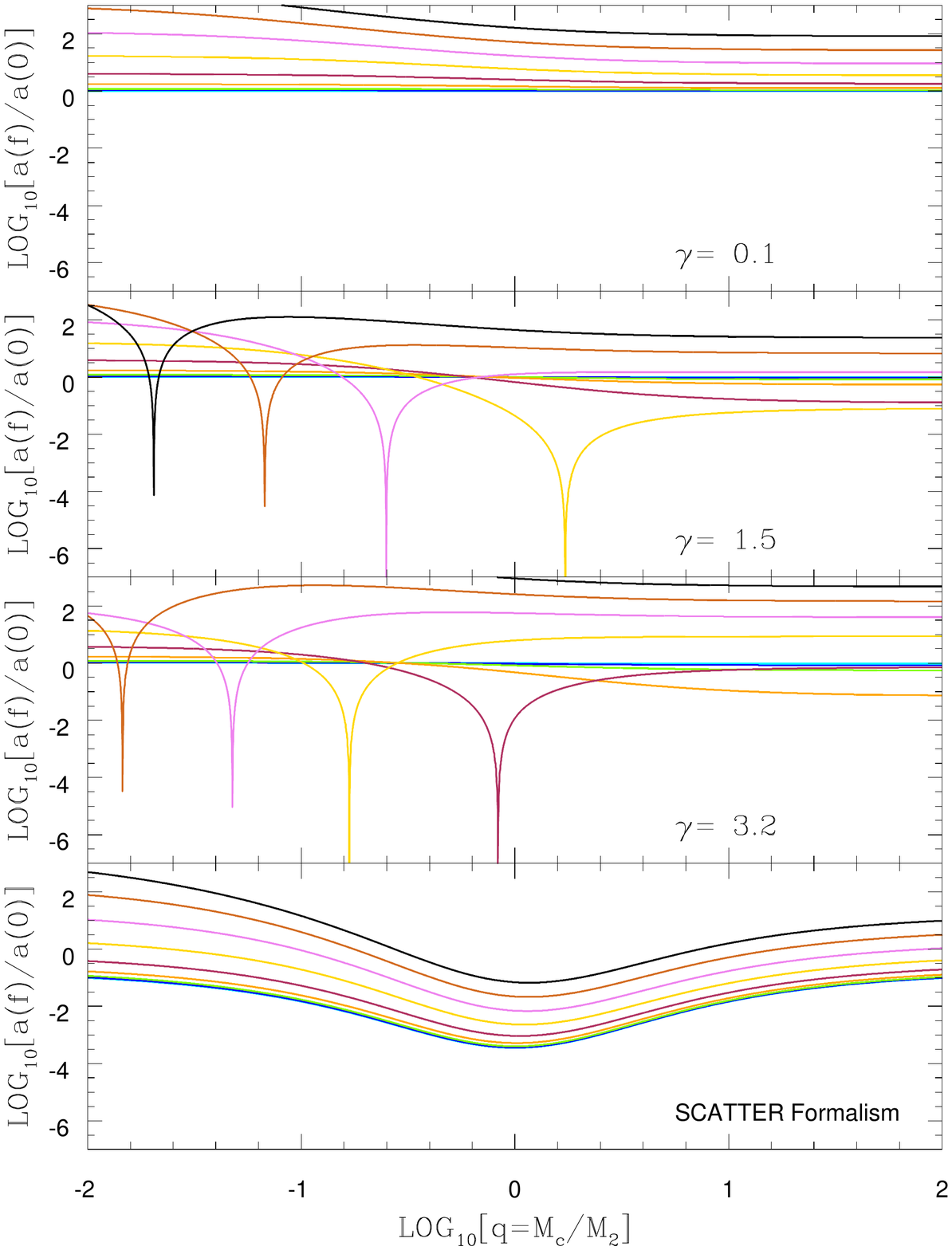}
    \caption{In both the $\gamma$ formalism and the new formalism, the ratio $a(f)/a(0)$ is shown as a function of the final mass ratio $q=M_\mathrm{c}/M_\mathrm{2}$. The only other variable is $q_\mathrm{ec}=M_\mathrm{e}/M_\mathrm{c}$, which is samples by the separate curves as shown in the bottom panel of Figure~\ref{fig:ce_shrink_eta}.}
    \label{fig:ce_shrink_gamma}
\end{center}
\end{figure}

Second, 
 the shrinkage is a monotonic function of $q$ in the $\alpha$ formalism, 
 and not in the SCATTER formalism, where the maximum amount of shrinkage occurs at 
 $q=1.$ In addition, we note that to conduct a full comparison with the 
 SCATTER formalism, a range of different values of $A$ and $B$ must be considered.
If we refer to Figure~\ref{fig:ce_shrink_eta}, it becomes clear that, while shrinkage in the $\alpha$ formalism always increases with increasing envelope mass, the same is not true
 of the SCATTER formalism, as we have already discussed. 
 
 Finally, we see that the $\alpha$ formalism generally produces more shrinkage. We might therefore expect it to yield more mergers within the CE, while fewer may occur in the SCATTER formalism. Thus, SCATTER may allow more systems to survive a first
 CE epoch. It may also lead to merger events that occur at later times. If  predictions of similar types are made by detailed BPS simulations, these comparisons will
 be testable.

\subsection{The \texorpdfstring{$\gamma$}{Gamma} Formalism}
\label{sec:compgamma}

In the $\gamma$ formalism \citep{nvyp00,nt05} the CE is ejected because of angular momentum imparted by the binary. It was invented to explain post CE binaries with a mild or even no orbital shrinkage. The $\gamma$ formalism has only one free parameter which describes the strength of the proportionality between angular momentum and mass change. The change of the orbital separation can be described by
\begin{equation}
    \label{eq:argam}
    \frac{a(0)}{a(f)} = \left[\frac{1}{1+q_\mathrm{ec}}\right]^2\,
    \left[\frac{1+q\, (1+q_\mathrm{ec})}{1+q}\right]\,
    \left[\frac{1+q\, (1+q_\mathrm{ec})}{1+q\,+q\,q_\mathrm{ec}\, (1-\gamma)}\right]^2 .
\end{equation}
As in the SCATTER formalism, this angular-momentum based approach allows some orbits to experience only modest shrinkage, or even expansion. On the other hand, there are sharply defined regions
of the parameter space that are associated with dramatic orbital shrinkage. In Figure~\ref{fig:ce_shrink_gamma} this fine tuning is visible by the deep dips which shift their position depending on the value of $\gamma$. The  SCATTER formalism does not exhibit such singular behavior.

In this regard the SCATTER formalism can reproduce the main required features of both existing formalisms. At the same time the equation, which describes the orbital change between the initial and final state remains on a similar complexity. This allows an easy usage in any kind of code which needs to deal with CE evolution.

\section{Conclusions}
\label{sec:conclusions}

The reason the CE plays such an important role in the study of binary systems is that a significant fraction of binaries experience a CE.  Most important, the CE end state determines whether a new phase of mass transfer or merger follows. Being able to predict CE outcomes is crucial to correct computations of many event rates, such as the rates of SNe~Ia or of gravitational mergers.
As we have seen in this work however, and in fact common to all CE formalisms, there are unavoidable uncertainties. Fortunately, the key question is not whether we can make an exact prediction for an individual system. Rather it is whether, when we start with a full population of binaries, do the results match observations of binary populations?  

The predictions of existing approaches to CE parameterization
 have not been able to match
the full complement of present-day observations.
It will be important to use SCATTER to make many sets of predictions and to check them against observations.
As with all formalisms, checks can be performed in astronomical contexts ranging from cataclysmic variables
to high-mass X-ray binaries.  The delay-time distributions for SNe~Ia and measures of chemical enhancement can both
provide checks, as can the rates of BH-BH, BH-NS, and NS-NS mergers. Although we don't know yet how well
SCATTER will fare in these tests, its approach has the advantage that, at its heart, is simple physics that we know applies. Furthermore the formalism is complex enough, in comparison with the standard $\gamma$ formalism, to be flexible. 

Our original motivation in developing SCATTER was to find a natural way to compute CE end states for hierarchical triples and other higher-order multiples. By focusing on the interaction of  ``Single Components'' with the CE,
one can develop natural generalizations in which there are more than two components. The fact that a 
simple functional form for the parameters can 
emerge from the formalism is a bonus associated with the formalism's use of angular momentum. 

Generalizations in addition to the application to higher-order multiples are also to be expected. Of most interest perhaps is the possibility of using more than one component of the angular momentum. The mathematical advantages that would provide are that computations for more complex systems may be possible. From a physical perspective, when the components of the binary (or higher-order multiple) can impart angular momentum in all three directions, the end-state 
can generally be expected to spin, leading to predictions that can be tested.

SCATTER suggests several directions for future research. One type of additional work
will be the application of the formalism to cases not considered in this paper.  The basic principles discussed in Section 3 apply to a wide range of systems, the phenomenological
approach we have employed allowed us to derive functional parameters for only the
types of systems identified as post-CE binaries in which a WD has a MS, WD, or NS companion.
Future investigations will consider
planetary systems and, at the other end of the mass spectrum, black holes.
We expect that this is just the first implementation of the SCATTER formalism and that its basic principles will prove useful in predicting the evolution of 
systems containing different types of stars, including those with high multiplicities.


\section*{Acknowledgements}

RD acknowledges support from the National Science Foundation, through NSF AST-2009520 and NSF PHY-1748958.
She also acknowledges useful discussions with Ilya Mandel and Ashley Ruiter. 
MK was partly supported by Grant No 12090040, and 12090043 of the Natural Science Foundation of China.
MK acknowledges support from the Swiss National Science Foundation Professorship Grant (PP00P2 211006; PI Fragos).
YG is funded by the Royal Society and the K.C. Wong Education Foundation.
CK acknowledges funding from the UK Science and Technology Facility Council (STFC) through grant ST/R000905/1 and ST/V000632/1, and a travel support from Z. Han.
This team collaboration started at a conference ``Progenitors of Type Ia Supernovae'' in Lijiang, 2019.
We thank the referee, Noam Soker, for helpful comments.



\bibliographystyle{aasjournal}
\bibliography{CE_refs.bib} 



\appendix
\section{The minimum and maximum envelope mass}
\label{sec:envelopemassrange}
\begin{figure}
    \centering
    \includegraphics[width=\columnwidth]{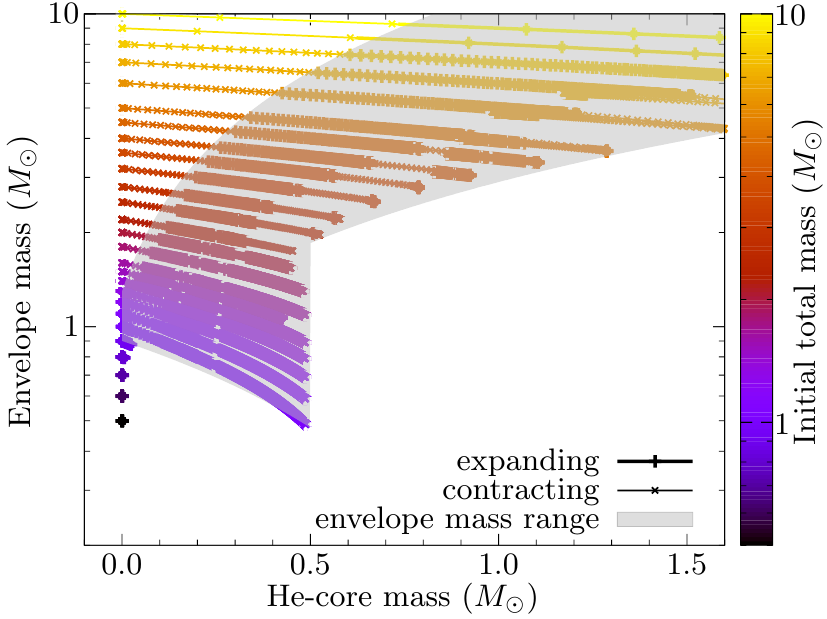}
    \caption{Helium core mass vs. envelope mass for stellar models with initial masses up to $\unit{10}{\Msun}$ and an age below the age of the Universe. Top: models calculated with BEC at $Z=0.0088$ \citep{ktl+18}.
    The colour shows the initial stellar mass. The think lines with the pluses are the phases of stellar expansion and the thin lines with the crosses are phases of contraction.}
    \label{fig:EnvelopeMassRange}
\end{figure}
\begin{figure*}
\centering
    \includegraphics[width=3.5 true in]{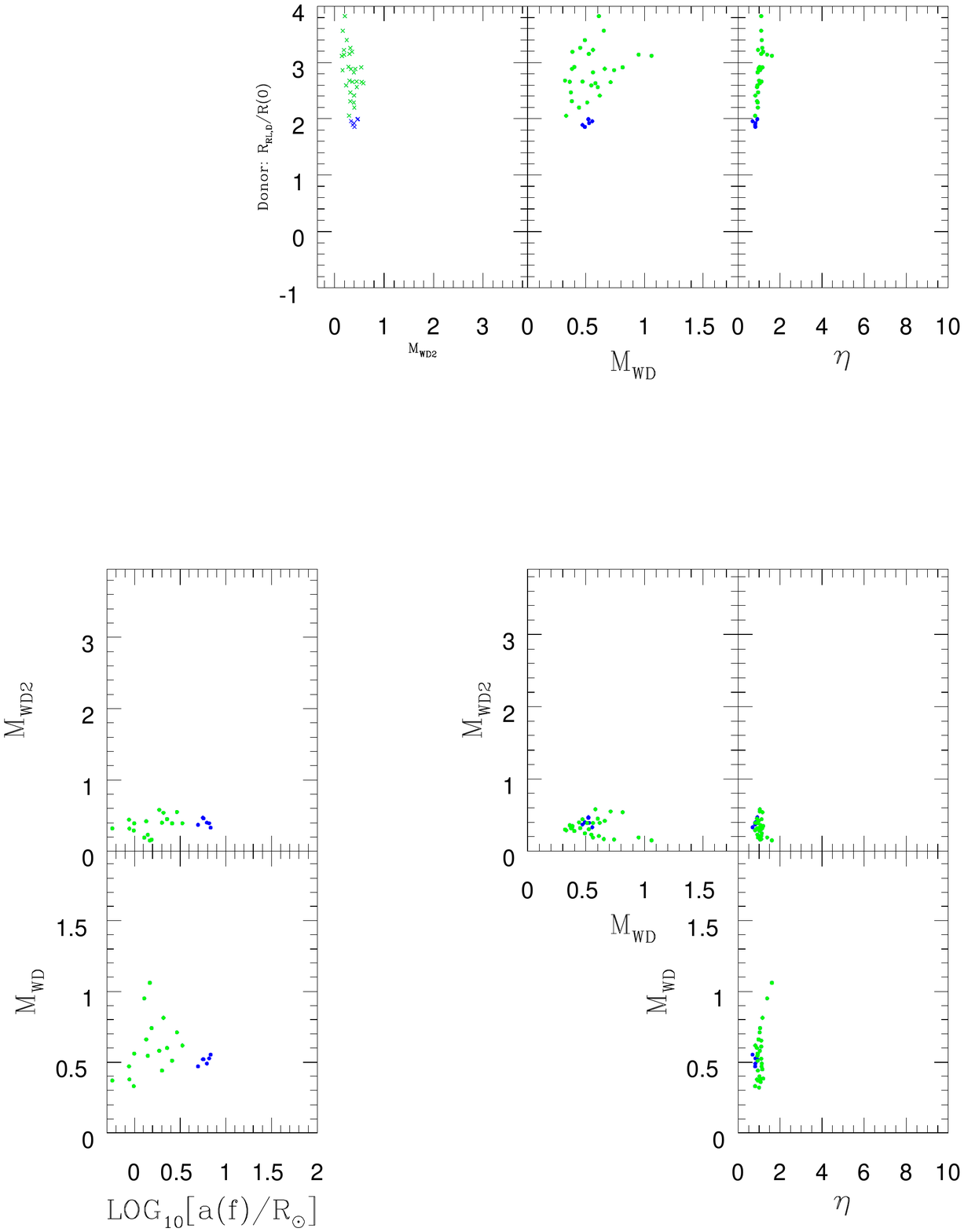}
    \includegraphics[width=3.5 true in]{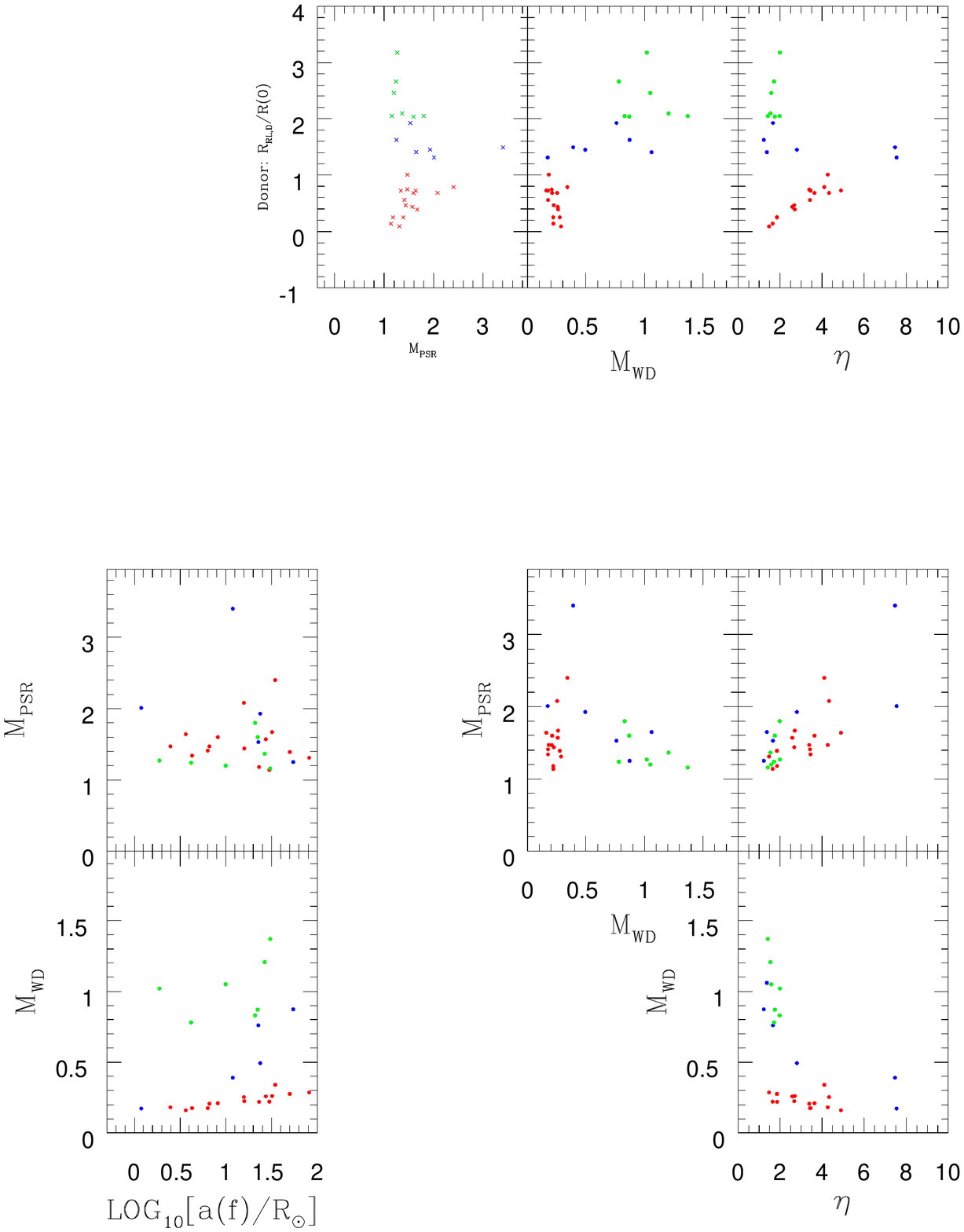}
    \caption{Like Figure~\ref{fig:ce_ms} but showing WD-WD binaries (left set of panels) and WD+NS binaries (right set of panels).}
    \label{fig:ce_wdns}
\end{figure*}

We have used the stellar models presented in \citet{ktl+18} (calculated with BEC, $Z=0.0088$)
to determine a minimum and a maximum mass for the envelope above the helium rich core. From those models we get two rough descriptions for the envelope mass range given a core mass, see Fig.~\ref{fig:EnvelopeMassRange}.
\begin{eqnarray}
    M_\mathrm{e}^\mathrm{min} &=&
    \begin{cases}
        \unit{0.9}{\Msun}-0.85\,M_\mathrm{c} & \text{if }M_\mathrm{c}\leq\unit{0.5}{\Msun},\\
        \unit{0.8}{\Msun}+2.1\,M_\mathrm{c} & \text{otherwise},
    \end{cases}\\
    M_\mathrm{e}^\mathrm{max} &=& \unit{1.3}{\Msun}+10.5\,M_\mathrm{c}.
\end{eqnarray}
When using the equations above we used additional limitations, e.g., by setting an additional solid limit on the minimum and maximum, see \S~\ref{sec:valueofeta}.

\section{Binaries with a White Dwarf and Compact Companion}
\label{sec:WDCO}
Table~1 indicates that we had both WD-WD and WD-NS post-CE candidate binaries.
We considered these in a  manner exactly analogous to the way we considered WD-MS
binaries. Figure B.1 shows for WD-WD binaries (top set of panels) and for WD-NS binaries (bottom set of panels) the same set of graphs as shown in Figure 3. The primary difference in these calculations is that Star~2 is a compact object. There is therefore no opportunity for a second epoch of mass transfer\footnote{If gravitational radiation brings the two compact objects close enough to each other that one fills its Roche lobe, these binaries could represent rare systems, such as AM CVn stars in which a compact object donates mass to compact companion.}. 

Unfortunately, double-compact binaries are dim and more challenging to identify than binaries consisting of a compact object in orbit with an extended star.  we therefore have fewer examples of them. Patterns easily discernible in the WD-MS systems are
more difficult to identify. For CE end states, the most important feature is that 
there is no single value, or limited range of values, of $\eta$ associated with these sets of binaries. Fortunately though, as we saw in Figure~\ref{fig:eta_2}, the WD-WD and WD-NS binaries are consistent with the same functional form we derived for WD-MS binaries. That is, $\eta$ can be written as $A\, \log_{10}[M_\mathrm{e}/M_\mathrm{t}] + B$.  

\end{document}